\documentclass[preprint,12pt]{elsarticle}

\usepackage{amssymb}
\usepackage{amsmath}
\usepackage{bm}
\usepackage{graphicx}
\usepackage{booktabs}
\usepackage{newtxtext}
\usepackage{newtxmath}
\usepackage{natbib}
\usepackage{hyperref}
\usepackage{subcaption}
\usepackage{siunitx}
\usepackage{cleveref}
\usepackage{float}
\hypersetup{
    colorlinks = true,
    urlcolor   = blue,
    citecolor  = blue,
}

\journal{International Journal of Multiphase Flow}

\begin{document}

\begin{frontmatter}

\title{Dispersion and clustering of deformable droplets in turbulence}

\author{Yushu Lin, John Palmore Jr}

\affiliation{organization={Department of Mechanical Engineering, University of Washington},
            addressline={3900 E Stevens Way NE}, 
            city={Seattle},
            postcode={98195}, 
            state={WA},
            country={USA}}

\begin{abstract}
Motivated by the application of spray combustion in the aviation industry, this work investigates the dispersion of non-spherical droplets in turbulence. The most common strategy for modeling sprays relies on the Lagrangian particle tracking (LPT) method, which represents the spray as a discrete collection of spherical particles. One limitation of this approach is that it neglects the influence of droplet deformation on the spray dynamics. Prior studies have highlighted the importance of non-sphericity in droplet vaporization, combustion and drag coefficient. However, these works are restricted to idealized configurations such as an isolated droplet in a uniform flow. To study droplet deformation in a more realistic configuration, we adopt homogeneous isotropic turbulence (HIT) as the framework to investigate its effect on droplet dispersion. {Droplets of various Stokes number are studied to investigate the interplay between deformation and inertia. Analysis of droplet statistics reveals that the impact of droplet deformation on both dispersion and clustering is dependent on the inertia regime. For weakly-inertial droplets, deformation weakens both dispersion and preferential concentration, whereas for strongly-inertial droplets, deformation tends to enhance preferential concentration while weakening dispersion. The results also suggest that to achieve the same level of clustering, deformed droplets require a higher Stokes number. Interestingly, for non-inertial droplets, the deformation seems to induce an effective inertia. This is verified by a comparison between the full unsteady Taylor Analogy Breakup (TAB) model and its steady-state limit, which suggests that unsteady shape dynamics affect temporal correlation statistics, but leave the mean clustering pattern unchanged. These findings demonstrate that accounting for droplet deformation and its unsteady shape oscillation is essential for accurately predicting droplet dispersion and clustering in turbulence.}
\end{abstract}

\begin{highlights}
\item Deformation suppresses dispersion while increasing clustering in strongly-inertial droplets.
\item Deformation suppresses both dispersion and clustering in weakly-inertial droplets.
\item Resolving the unsteady effects of deformation is important to predicting droplet dispersion.
\end{highlights}


\end{frontmatter}

\section{Introduction}
\label{sec:introduction}
Droplets appear in a wide range of scenarios in nature and industry, such as spray combustion in gas turbine engines, sea spray generation by wave breaking, and volcanic ash dispersion during eruption. This work is motivated by spray combustion in which a liquid fuel is injected into an engine where it atomizes into droplets. The rates of combustion and energy release in the engine are limited by droplet dispersion and the evaporation rate. In particular, this work investigates the dynamics of droplet motion in turbulence focusing on understanding the importance of droplet deformation due to flow shear on the dispersion and clustering behaviors of droplets. 

The theory of droplet dispersion was pioneered by \citet{taylorDiffusionContinuousMovements1922}. Taylor conducted the first analytical investigation on determining diffusion coefficient of a continuous passive scalar from the Lagrangian velocity autocorrelation. Based on Taylor's foundational work, \citet{batchelorDiffusionFieldHomogeneous1949} developed a comprehensive three-dimensional Eulerian framework to describe the diffusion of fluid particles in turbulence. \citet{tchenMeanValueCorrelation1947} was the first to extend Taylor's work to small discrete solid particles. Tchen assumed that particles faithfully follow flow streamlines, in which case particle dispersion will be identical to the fluid dispersion. \citet{gouesbetDispersionDiscreteParticles1984} later generalized Tchen's one-dimensional theory into a three-dimensional dispersion tensor. Using a reduced dispersion coefficient defined as the ratio of particle to fluid dispersion, Gouesbet et al. showed that while dense particles can initially disperse faster than fluid particles, over longer times fluid dispersion always surpasses that of inertial particles.

Subsequent work focused on how particle dispersion deviates from fluid dispersion due to the effect of inertia and gravity. \citet{friedlanderBehaviorSuspendedParticles1957} showed that particle inertia leads to a reduction in dispersion at short times, and converges to fluid dispersion at sufficiently long times. Later, a key phenomenon called the crossing trajectory effect was identified by \citet{yudinePhysicalConsiderationsHeavyParticle1959} when a steady drift is present due to gravity. The drift causes heavy particles to continuously cross different fluid eddies, reducing particle dispersion relative to fluid dispersion. This effect was further quantified by \citet{nirEffectSteadyDrift1979} and explored in theoretical and numerical studies for particles with both finite inertia and gravitational settling \citep{reeksDispersionSmallParticles1977,wangDispersionHeavyParticles1993,squiresMeasurementsParticleDispersion1991,bankoParticleDispersionPreferential2023}.

It was further recognized that particle dispersion is influenced by the coherent structures of turbulence \citep{croweParticleDispersionCoherent1985, goreParticleDispersionLarge1989}. \citet{maxeyGravitationalSettlingAerosol1987} provided a theoretical derivation demonstrating that inertial particles tend to accumulate in high-strain or low-vorticity regions. The first numerical evidence was reported by \citet{squiresPreferentialConcentrationParticles1991}, who observed this clustering phenomenon in direct numerical simulations and termed it as preferential concentration. Later, \citet{eatonPreferentialConcentrationParticles1994} and \citet{balachandarTurbulentDispersedMultiphase2010} reviewed the underlying mechanisms of preferential concentration along with related experimental and numerical studies.

{For droplet and particulate combustion, this non-uniformity in dispersion is relevant as it affects the vaporization rate \citep{chenDNSStudyPulverized2023}. Generally, higher concentrations of droplets have lower evaporation rates and \textit{vice versa}. Accordingly, measurement of particle clustering is an implicit reflection of evaporation. A simple and widely used method to measure clustering is through tracking variations in the local droplet volume fraction \citep{mirandaHighStokesNumber2020}. However, this approach is only meaningful when the droplet volume fraction is sufficiently large. For dilute flows, a Voronoi analysis approach was popularized by \citet{monchauxPreferentialConcentrationHeavy2010}.}

Most of the studies on particle dispersion in turbulence appear to be for spherical particles. For non-spherical particles, a significant portion of the fundamental research literature focuses on particle orientation and rotation dynamics in turbulent flows.  \citet{marchioliOrientationDistributionDeposition2010}, \citet{siewertOrientationStatisticsSettling2014}, and \citet{njobuenwuDynamicsSingleNonspherical2015} examined the orientation and translation of ellipsoidal particles in turbulent channel flow and decaying isotropic turbulence. \citet{jiePreferentialOrientationTracer2019} further showed that preferential orientation exists even for inertialess spheroids. \citet{giahiFullyResolvedSimulation2024} performed fully resolved simulations of spherical and non-spherical particles in turbulent channel flow, showing that shape significantly alters particle trajectories, rotation, and wake structures, demonstrating that a simple drag correction based on the equivalent sphere assumption is insufficient. The review by \citet{vothAnisotropicParticlesTurbulence2017} summarized anisotropic particle orientation and rotation dynamics in turbulence.

The significant impact of non-sphericity on bulk particle motion has also been recognized and has been investigated by researchers. \citet{brennerTaylorDispersionSystems1979,brennerTaylorDispersionSystems1981} investigated non-spherical Brownian particles, and found that particle dispersion is proportional to the diameter. Experimental work by \citet{dubeDynamicsNonsphericalParticles2013} on non-spherical particles in rotating drums revealed that non-spherical particles have lower axial dispersion than spherical ones due to preferential orientation. \citet{saxbyImpactParticleShape2018} modeled atmospheric volcanic ash dispersion, and observed that non-spherical particles travel further than spherical particles, and the sensitivity of dispersion to particle shape is dependent on particle diameter. \citet{elshorbagyEffectSolidParticle2022} performed simulations of non-spherical particles in a cyclone separator, and the results show higher tangential velocities of non-spherical particles in rotating flows. \citet{yangMixingDispersionBehaviours2022} explored the mixing and dispersion behavior of non-spherical particles in a bubbling fluidized bed, and concluded that particle dispersion is a non-linear function of aspect ratio, peaking at around 0.75. {However, these investigations are for either non-turbulent flows or for specific scientific and engineering configurations. Fundamental studies in canonical turbulence configurations should be performed to characterize the importance of these effects in more general contexts.}

A common feature across these studies is that they focused on rigid non-spherical particles with fixed shapes. Consequently, the central question becomes how particles orient relative to the local flow gradient. For liquid droplets, this framework is not directly applicable, as droplet shape changes dynamically under the competing influences of aerodynamic stress and surface tension, resulting in unsteady oscillations absent in rigid particles. The primary effect of non-sphericity in droplets therefore does not originate from the orientation but from the instantaneous shape dynamics and the resulting modification of drag. Recent work by \citet{linNonsphericalDropletDispersion2024} on non-spherical droplets in homogeneous isotropic turbulence has begun to explore this aspect. However, that study only considered a single combination of droplet diameter and density, leaving the effects of inertia and variations in droplet deformation underexplored. Furthermore, it did not account for transient droplet deformation, which the authors noted as physically inaccurate due to the relatively large capillary time of the droplets.

When extending the analysis from solid particles to liquid droplets, the physical complexity increases significantly due to the effect of droplet deformation. For solid particles with different fixed geometries, drag coefficient varies with shape. For example, \citet{cliftBubblesDropsParticles1992} reported oblate particles exhibit lower drag than rigid spheres at Reynolds number $Re<37$ but higher drag at $Re>37$. {For liquid droplets, however, droplet shape deforms dynamically due to the capillarity effect, which provides a restoring surface tension force to resist the deforming aerodynamic force. The competition between the two forces gives rise to unsteady shape oscillation that is absent in rigid particles, as described by the Taylor Analogy Breakup (TAB) model \citep{orourkeTabMethodNumerical1987}.} In addition to the change of droplet shape, the viscous droplet interface couples the gaseous and liquid phases, inducing an internal circulation that forms a boundary layer at the interface and alters the overall drag. Early studies on viscous spherical droplets, such as \citet{harperMotionSphericalLiquid1968}, modeled the internal circulation as Hill’s vortex at high Reynolds numbers, while \citet{rivkindFlowStructureMotion1977} solved the axisymmetric stream-function and vorticity equations to capture internal circulation at intermediate Reynolds numbers. Further, \citet{fengDeformableLiquidDrop2010} introduced a two-layer concept to capture the boundary layer at droplet interface and established a drag correlation for viscous spherical droplets at intermediate viscosity ratios. \citet{linNumericalStrategyInvestigating2022} performed DNS to quantify internal circulation of deformed droplets. The work showed that higher pressure strengthens circulation while larger deformation accelerates its decay, and that circulation can be quantified via the liquid-to-gas density ratio. More recently, \citet{niDeformationBreakupBubbles2024} provided a comprehensive review, showing that deformation fundamentally alters hydrodynamic forces, leading to drag reduction, turbulence modulation, and enhanced interfacial heat and mass transfer.

For studies on drag coefficient of deformed droplets, \citet{haywoodNumericalSolutionDeforming1994} employed a finite-volume non-orthogonal adaptive-grid system to study evaporating deformed droplets; \citet{helenbrookQuasisteadyDeformationDrag2002} examined the combined effects of deformation and internal circulation, identifying oblate, prolate, and dimpled shapes and proposing an aspect-ratio correlation for spheroidal droplets; \citet{lothQuasisteadyShapeDrag2008} comprehensively reviewed drag coefficient of different particle types, including solid particles, viscous spherical drops/bubbles and deformed drops/bubbles, and established a unified drag correlation. More recently, \citet{linEffectDropletDeformation2022} used direct numerical simulations to investigate the coupled effects of deformation and internal circulation on droplet drag at high pressure, finding that the drag coefficient increases with increasing Weber number and decreasing liquid-to-gas density ratio, and deriving an analytical expression to explain these trends; \citet{gugliettaDeformationEllipsoidalDroplets2025} used more complex droplet deformation models to assess model accuracy, and it was suggested that simple models are sufficient for slightly deformed droplets, while more advanced ellipsoidal models are required for highly deformed droplets.

In this work, we employ numerical approaches to investigate how droplet dispersion and clustering are affected by droplet deformation. An in-house code developed for simulating multiphase flows is adopted to simulate polydisperse droplets at different diameters in turbulence. Considering that the homogeneous isotropic turbulence is a canonical framework to study turbulence in a simplified and well-controlled manner, it is adopted in our study as the carrier phase. The carrier phase is modeled by using an in-house solver developed by \citet{palmoreTechniqueForcingHigh2018} for solving high Reynolds number turbulent flows. Droplets are modeled by using a Lagrangian particle tracking (LPT) solver designed to simulate the full range of dilute-to-dense particle laden flows \citep{capecelatroEulerLagrangeStrategy2013}. The unsteady deformation of droplets is modeled by the TAB model proposed by \citet{orourkeTabMethodNumerical1987}, and the drag coefficient of droplets is modeled by the Dynamic Drag Model (DDM) proposed by \citet{liuModelingEffectsDrop1993}. Our analysis of droplet dispersion compares the parameters developed by \citet{taylorDiffusionContinuousMovements1922}, including the droplet velocity autocorrelation, droplet Lagrangian integral time scale, and droplet dispersion coefficient. We also develop a local vortex coordinate system to decompose droplet velocity to reveal the interaction between droplets and the vortical structures of the flow. Furthermore, droplet clustering is analyzed using the Voronoi approach \citep{monchauxPreferentialConcentrationHeavy2010}.

The remainder of the paper is organized as follows. We begin in \Cref{sec:methods} with a detailed description of numerical methods used in this study, including the flow solver for the carrier phase in \Cref{sec:sub:fs} and the LPT droplet model for solving droplets in \Cref{sec:sub:ds}. \Cref{sec:measurements} introduces methods used for quantifying droplet dispersion and clustering in turbulence. In \Cref{sec:results} we present results on droplet dispersion and clustering by using methods introduced in \Cref{sec:measurements}. Finally in \Cref{sec:conclud} we offer conclusions on how droplet dispersion and clustering are affected by droplet deformation for advanced spray modeling in turbulent combustion applications.

\section{Numerical methods}
An Eulerian–Lagrangian framework is adopted to perform numerical simulations. The carrier fluid phase is solved on an Eulerian mesh, while the dispersed droplets are tracked as Lagrangian point particles. Details of the carrier phase and dispersed phase solvers are discussed in \Cref{sec:sub:fs} and \Cref{sec:sub:ds}.
\label{sec:methods}
\subsection{Problem Formulation}
\begin{figure}[H]
    \centering
    \includegraphics[width=0.6\linewidth]{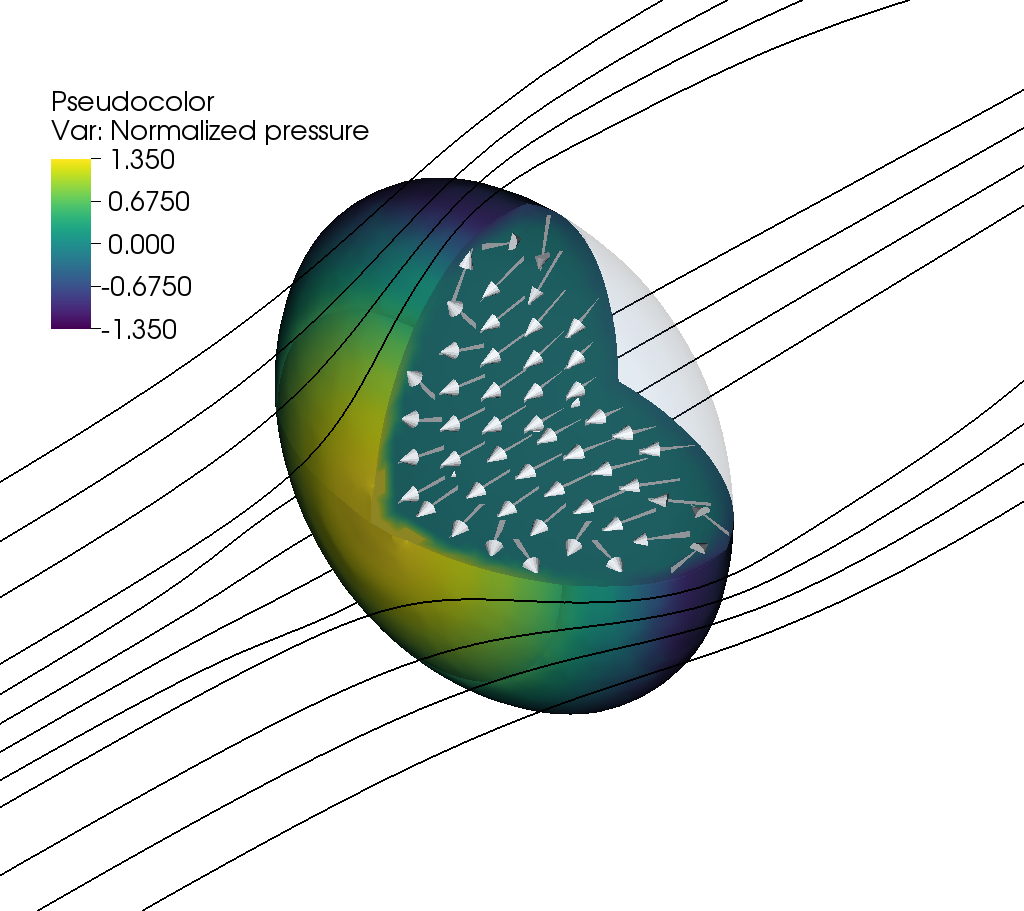}
    \caption{Highly deformed droplet with internal circulation obtained from three-dimensional simulation. Original image from authors. Also used as graphical abstract for \citep{linEffectDropletDeformation2022}.}
    \label{fig:deformation}
\end{figure}

We simulate forced homogeneous isotropic turbulence in a cubic box with periodic boundary conditions, and add one-way coupled droplets to the dispersed phase. The controlling parameters of the problem are the Stokes number and Weber number, defined as
\begin{equation}
    \mathrm{St}=\frac{\tau_\mathrm{d}}{\tau_\eta},\ {\mathrm{We}=\frac{\rho_\mathrm{f}{\Delta{\bm{u}}}^2d}{\sigma}},
    \label{eqn:def}
\end{equation}
where $\tau_\mathrm{d}$ is droplet relaxation time defined as $\tau_\mathrm{d}=\frac{\rho_\mathrm{d}}{\rho_\mathrm{f}}\frac{4}{3}\frac{d}{C_\mathrm{D}}\frac{1}{\left|\Delta\bm{u}\right|}$, $\tau_\eta$ is Kolmogorov time scale of turbulence, {$\Delta{\bm{u}}=\bm{u}_\mathrm{d}-\bm{u}_\mathrm{f}\left(\bm{x}_\mathrm{d}\right)$} is the relative velocity between droplets and the flow, $d$ is droplet diameter, $u_\mathrm{d}$ is droplet velocity, and $\sigma$ is droplet surface tension. The Stokes number reflects the inertia of droplets, while the Weber number reflects the deformation of droplets. \Cref{fig:deformation} depicts a highly deformed droplet at high $\mathrm{We}$ \citep{linEffectDropletDeformation2022}. Due to the relatively high aerodynamics stresses, the droplet deforms into an oblate spheroidal shape. This changes the streamlines of the flow around the droplet and alters the drag profile. A secondary flow is generated inside the droplet which further affects the droplet drag.

\subsection{Flow Solver}
\label{sec:sub:fs}
The fluid phase is resolved by using an in-house code developed by \citet{palmoreTechniqueForcingHigh2018} for direct numerical simulation of high Reynolds number turbulent flow in a Eulerian framework, and modified for particle-laden flow by \citet{mirandaHighStokesNumber2020}. The flow is governed by the Navier-Stokes equations
\begin{equation}
    \frac{\partial\left(\rho\bm{u}\right)}{\partial t}+\nabla\cdot\left(\rho\bm{u}\otimes\bm{u}\right)=-\nabla p+\nabla\left[\mu\left(\nabla\bm{u}+\left(\nabla\bm{u}\right)^T\right)\right]+\bm{f}^\mathrm{S}_\mathrm{d}+\bm{f},
    \label{momentum}
\end{equation}
where $\bm{f}_\mathrm{d}^\mathrm{S}$ represents the momentum exchange between fluid and particles. {Because the current study focuses on how turbulence disperses droplets, we isolate the effect of turbulence on deformed droplets by using one-way coupling droplet-laden flow simulations, i.e. $\bm{f}_\mathrm{d}^S=0$. The volume fraction of droplets in our simulations is on the order of $10^{-8}$, which is far below the threshold for the one-way coupling regime identified by \citet{elghobashiParticleladenTurbulentFlows1991}. Although high-$\mathrm{St}$ droplets may have a relatively large reaction force on the flow, the small mass loading (around 0.259\% at most) of droplets suggests that the overall effect of the dispersed phase on the flow is still small.}

The final term, $\bm{f}$, is an external forcing used to maintain the statistical steadiness of turbulence. To balance the decay of turbulent kinetic energy due to viscous dissipation, a filtered linear forcing strategy developed by \citet{palmoreTechniqueForcingHigh2018} is adopted for $\bm{f}$. The forcing term \Cref{forcing} is applied in physical space and is proportional to a low-pass filtered velocity field for linear forcing,
\begin{equation}
    \bm{f}=\rho A\hat{\bm{u}},
    \label{forcing}
\end{equation}
where $A$ is an arbitrary constant, and $\hat{\bm{u}}$ is the filtered velocity field. $A$ is chosen such that the forcing term will enforce turbulence to maintain its initial properties and become statistically steady. The procedure for determining $A$ is referenced to \citet{mirandaHighStokesNumber2020}.

\subsection{Droplet Solver}
\label{sec:sub:ds}
The initial spatial distribution and velocity of the droplets are obtained from a precursor simulation, where droplets are initialized with zero velocity and a uniform distribution. The droplets are tracked using the Lagrangian particle tracking method. Each droplet is treated as a Lagrangian particle and is governed by
\begin{equation}
    \frac{d\bm{x}_\mathrm{d}}{dt}=\bm{u}_\mathrm{d},
    \label{lpt_u}
\end{equation}
\begin{equation}
    m\frac{d\bm{u}_\mathrm{d}}{dt}=\bm{F}_\mathrm{d}^\mathrm{S},
    \label{lpt_a}
\end{equation}
where $\bm{F}^\mathrm{S}_\mathrm{d}$ is the surface force acting on the droplet. In our simulations, $\bm{F}^S_\mathrm{d}$ is the drag force of droplets. Due to the low volume fraction, droplet-to-droplet collisions are not considered.

{In this study, we employ the dynamic drag coefficient correlation proposed by \citet{liuModelingEffectsDrop1993}, which accounts for the correction of droplet shape change. The shape of deformed droplets is assumed to be oblate spheroids, and the equatorial plane of the oblate droplets is perpendicular to the local flow direction. Then the DDM drag correlation is}
\begin{equation}
    {C_\mathrm{D}=C_{\mathrm{D},\mathrm{sphere}}\left(1+2.632y\right),}
    \label{Cd_DDM}
\end{equation}
{where $C_{\mathrm{D},\mathrm{sphere}}$ is the drag correlation of rigid spheres given by}
\begin{equation}
    {C_{\mathrm{D},\mathrm{sphere}}=\frac{24}{Re_\mathrm{d}}\left(1+\frac{1}{6}Re_\mathrm{d}^{2/3}\right),}
    \label{Cd_sphere}
\end{equation}
{$Re_\mathrm{d}$ is droplet Reynolds number, and $y$ in \Cref{Cd_DDM} is the non-dimensionalized distortion of droplets defined as $y=2\Delta r/r$, where $r$ is original droplet radius and $\Delta r$ is the displacement of the droplet equator from its equilibrium position. 
$y$ is obtained by solving the equation of droplet distortion given by TAB model proposed by \citet{orourkeTabMethodNumerical1987} in \Cref{TAB},}
\begin{equation}
    \ddot{y}=\frac{2\rho_\mathrm{f}}{3\rho_\mathrm{d}}\frac{u^2}{r^2}-\frac{8\sigma}{\rho_\mathrm{d}r^3}y-\frac{5\mu_\mathrm{d}}{\rho_\mathrm{d}r^2}\dot{y}.
    \label{TAB}
\end{equation}

\section{Measurements of Droplet Motion}
\label{sec:measurements}
\subsection{Quantification of droplet dispersion}
\label{sec:sub:disp_coeff}
The mean-square displacement $\langle X^2 \rangle $ is representative to quantify droplet dispersion in homogeneous isotropic turbulence. A more commonly used parameter is the time derivative of $\langle X^2 \rangle $, named as dispersion coefficient $K=\frac{d}{dt}\langle X^2 \rangle $, which represents the statistical rate at which droplets are conveyed away from their original positions.

The pioneering work by \citet{taylorDiffusionContinuousMovements1922} demonstrated that for homogeneous isotropic turbulence, dispersion can be fully characterized from the autocorrelation of Lagrangian particle velocity, $R_{ij}$, as defined in \Cref{eqn:Rij}, where $i,j$ refers to the directions of droplet velocity. 
\begin{equation}
    R_{ij}\left(t\right)=\frac{\langle {u}_{\mathrm{d},i}\left(t_0\right){u}_{\mathrm{d},j}\left(t_0+t\right) \rangle }{\sqrt{\langle u_{\mathrm{d},i}^2\left(t_0\right) \rangle }\sqrt{\langle u_{\mathrm{d},j}^2\left(t_0\right) \rangle }}
    \label{eqn:Rij}
\end{equation}

The mean-square displacement of a particle within a certain time interval can be calculated from the autocorrelation, as shown in \Cref{eqn:XX}, and similarly for the dispersion coefficient $K$ using \Cref{eqn:K}. 
\begin{equation}
    \langle {X}_{i}^2 \rangle ={2{\langle {{u}}_{\mathrm{d},i}^2 \rangle }\int_0^t \int_0^\tau R_{ii}\left(\xi\right) \,d\xi d\tau}
    \label{eqn:XX}
\end{equation}
\begin{equation}
    K\left(t\right)=\frac{d}{dt}\langle {X}^2 \rangle ={\frac{2}{3}\sum_i {\langle u^2_{\mathrm{d},i}\left(t\right) \rangle } \int_0^t R_{ii}\left(\xi\right) \,d\xi}
    \label{eqn:K}
\end{equation}

Another important parameter is the Lagrangian time, $\tau_L$. It is a characteristic dispersion time defined in terms of the autocorrelation, and it represents the time for droplet motion to decorrelate from its initial position.
\begin{equation}
    \tau_{L}=\sum_i{\int_0^\infty R_{ii}\left(t\right) \, dt},
    \label{eqn:t_L}
\end{equation}

 It distinguishes the behavior of droplet dispersion under two distinct asymptotic regimes, derived by \citet{taylorDiffusionContinuousMovements1922}. When $t\ll\tau_\mathrm{L}$, droplet velocity autocorrelation does not significantly drop from unity, and the mean-square displacement of droplets is under the ballistic regime and is proportional to $t^2$. When $t\gg\tau_\mathrm{L}$, the mean-square displacement transitions to the diffusive regime and grows linearly with time. At sufficiently long times where $t\gg\tau_\mathrm{L}$, \Cref{eqn:K} reduces to \Cref{eqn:K_inf}, where $\tau_\mathrm{L}$ is the Lagrangian time scale defined in \Cref{eqn:t_L}. 
\begin{equation}
    K=2\langle u^2_\mathrm{d} \rangle \tau_\mathrm{L}
    \label{eqn:K_inf}
\end{equation}

A final measure of dispersion is the particle-to-fluid relative dispersion. \citet{gouesbetDispersionDiscreteParticles1984} defined the reduced dispersion coefficient $S$ to measure how the particles dispersion deviates from the fluid flow,
\begin{equation}
    S=\frac{K\left(t\right)}{K_\mathrm{f}\left(t\right)},
    \label{eqn:S}
\end{equation}
\begin{equation}
    S_{\infty}=\frac{K\left(t\right)}{K_\mathrm{f}\left(t\rightarrow\infty\right)},
    \label{eqn:S_inf}
\end{equation}
where $K$ is the droplet dispersion coefficient of droplets, and $K_\mathrm{f}$ is the dispersion coefficient of fluid particles. In this study, we use the dispersion coefficient of tracer particles as an estimate of $K_\mathrm{f}$.

{$S$ represents the instantaneous comparison between the particle dispersion and the fluid particle diffusion, while $S_{\infty}$ represents the asymptotic behavior of particle dispersion and determines how fast the dispersion occurs. If $S>1$, it means particles are dispersing more rapidly than fluid particles; if $S<1$, then it means particles are lagging behind the flow. $S_\infty$ determines the eventual dispersion of particles compared to the fluid.}

\subsection{Quantification of Preferential Concentration}
In turbulent flow, the interaction between droplets and vortices is an essential factor that alters the spatial distribution of droplets. The Stokes number quantifies the competition of inertia between droplets and the fluid. When $\mathrm{St}\ll1$, droplets behave like tracer particles that faithfully follow fluid streamlines. Conversely, if $\mathrm{St}\gg1$, droplets are largely unaffected by the flow, but maintain their own trajectories. For intermediate $\mathrm{St}\approx1$, droplets tend to cluster in regions of low vorticity and high strain rate. In particular, the preferential concentration phenomenon predicts that droplets will centrifuge out of vortices, \emph{i.e.} preferentially move away from the center of vortices \citep{balachandarTurbulentDispersedMultiphase2010}. {Based upon these physical regimes, this article will use the terms non-inertial ($St\ll1$), weakly-inertial ($St\sim1$), and strongly-inertial ($St\gg1$).}
\subsubsection{Velocity decomposition in local vortex coordinate}
\label{sec:sub:moffatt}
\begin{figure}[H]
    \centering
    \includegraphics[width=0.3\linewidth]{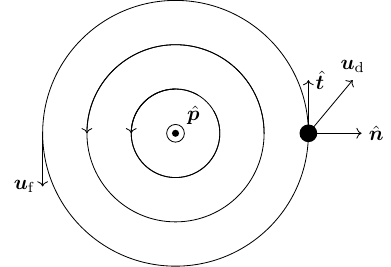}
    \caption{local vortex coordinate system; the sign $\odot$ at the center represents $\hat{\bm{p}}$ pointing out of the plane toward reader}
    \label{fig:coordinate}
\end{figure}

To better investigate the effect of droplet-vortex interaction, we introduce a local vortex coordinate system motivated by the kinematics of a line vortex. For an ideal line vortex, the flow can be decomposed into a swirling motion, a translational motion parallel to the vorticity direction, and a translational motion perpendicular to the vortex line. Such decomposition of the flow field intuitively illustrates the motion of the flow relative to the vortex line. In our case, to describe droplet motion relative to local vortical structures in a complex turbulent flow, we generalize this method by decomposing relative velocity $\Delta{\bm{u}}$ into three directions defined by three orthogonal unit vectors $\hat{\bm{p}},\hat{\bm{t}},\hat{\bm{n}}$, corresponding to directions: 1) parallel to local vortex, 2) tangent to local streamline, 3) and normal to local vortex, as illustrated in \Cref{fig:coordinate}. The unit vectors are defined as
\begin{align}
    \hat{\bm{p}}=\frac{\bm{\omega}}{\left|\bm{\omega}\right|},\\
    \hat{\bm{t}}=\frac{\bm{u}_\mathrm{f}}{\left|\bm{u}_\mathrm{f}\right|},      \\
    \hat{\bm{n}}=\hat{\bm{t}}\times\hat{\bm{p}},
\end{align}
where $\bm{\omega}=\nabla\times\bm{u}_\mathrm{f}$. {These velocity components are then normalized by the magnitude of $\Delta{\bm{u}}$, yielding the direction cosines $\cos\theta_{\mathrm{rel,p}}$, $\cos\theta_{\mathrm{rel,t}}$ and $\cos\theta_{\mathrm{rel,n}}$ of $\Delta{\bm{u}}$ with respect to $\hat{\bm{p}}$, $\hat{\bm{t}}$, and $\hat{\bm{n}}$. In the remainder of the paper, we refer to droplet velocity components in these three directions as $p$-, $t$- and $n$-components respectively. The choice of such coordinate helps determine whether or not a droplet is moving toward or away from a vortex depending only on the sign of $n$-component velocity.} 

\subsubsection{Voronoi analysis}
\label{sec:clustering}
{Measurement of $n$-component velocity only implies that preferential motion occurs, it does not quantify the amount of this motion. To quantify the amount of preferential motion, we use Voronoi analysis developed by \citet{monchauxPreferentialConcentrationHeavy2010}. A Voronoi cell is defined as a region surrounding a specific ``seed'' point, such that any other points within the region is closer to any other seed points. In our case, the seed point is the droplet. It is intuitive that a larger Voronoi cell corresponds to a region of lower particle concentration. Thus, the inverse of Voronoi cell volume represents the local concentration of particles. By comparing the probability density function of Voronoi cell volume and its standard deviation across different cases, we can effectively quantify the strength of droplet clustering.}

\section{Results}
\label{sec:results}
\begin{figure}[H]
    \centering
    \includegraphics[width=0.5\linewidth]{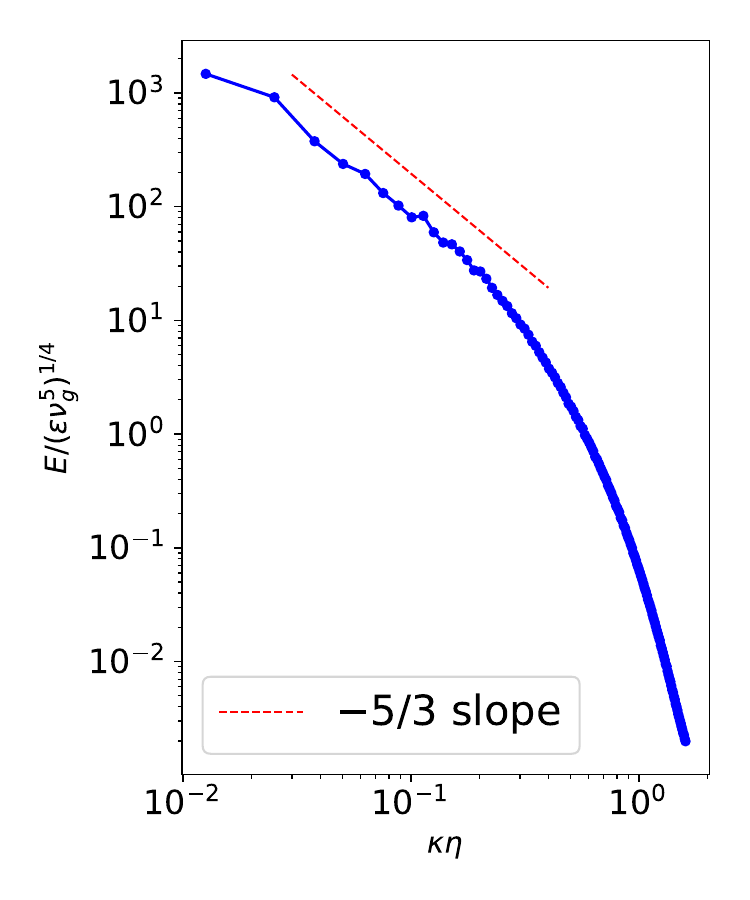}
    \caption{Energy spectrum of the HIT flow.}
    \label{fig:spectrum}
\end{figure}

We solve the turbulent flow in a cubic domain with side length $L=\SI{0.044}{m}$. The computational domain is discretized into $N=256$ cells along each direction. Parameters of the flow are listed in \Cref{tab:HIT}. In the table, $\mathrm{rms}\left(u\right)$ refers to the root-mean-square fluid velocity, $\nu$ is fluid viscosity, and $\eta$ is the Kolmogorov length scale which characterizes the smallest eddy size in turbulence. {The ratio of the Kolmogorov length scale to the grid spacing is $\frac{\eta}{\Delta}\approx0.462$ where $\Delta=L/N$, indicating that the DNS is resolved \citep{Pope_2000}. Furthermore, \Cref{fig:spectrum} shows the energy spectrum exhibits a distinct inertial subrange, as seen by the region with -5/3 slope. This verifies the physical validity of our homogeneous isotropic turbulence simulation.} $Re_{\lambda}$ refers to the Taylor-microscale based Reynolds number, and $\kappa_\Delta$ is the largest resolvable wavenumber on the grid.

\begin{table}[H]
    \centering
    \begin{tabular}{|c|c|c|c|c|c|c|}
    \hline
    $N$ & $\mathrm{rms}\left(u\right)$ ($\unit{m/s}$) & $\nu$ ($\unit{m^2/s}$) & $\eta$ ($\unit{m}$) & $Re_\lambda$ & $\kappa_\Delta\eta$ \\
    \hline
    $256$ & $0.3999$ & $6.024\times10^{-6}$ & $7.939\times10^{-5}$ & $156$ & $1.451$ \\
    \hline
    \end{tabular}
    \caption{Turbulent flow parameters}
    \label{tab:HIT}
\end{table}

To systematically investigate the coupled effects of inertia and deformation, the droplet density, viscosity, and surface tension are independently varied to control the Stokes and Weber numbers, while Ohnesorge number is fixed at $\mathrm{Oh}=0.1$. Droplet density varies between $\rho_\mathrm{d}=\SI{683.85}{kg/m^3}$ and $\SI{68385}{kg/m^3}$ across three different diameters at $d_0=\SI{1.567e-06}{m}$, $d_1=\SI{4.957e-06}{m}$ and $d_2=\SI{1.108e-05}{m}$. The controlling parameters of droplets are listed in \Cref{tab:drop}, where $\mathrm{St}^*$ is the expected Stokes number calculated based on Stokes' drag $\mathrm{St}^*=\frac{\rho_\mathrm{d}d^2}{18\mu_\mathrm{f}}\frac{1}{\tau_\eta}$, and $\mathrm{We}^*$ is the expected Weber number estimated based on \citet{abrahamsonCollisionRatesSmall1975a}'s theory that calculates droplet velocity by using \Cref{eqn:ud2}.
\begin{equation}
    \frac{\langle u_\mathrm{d}^2 \rangle }{\langle u_\mathrm{f}^2 \rangle }=\frac{1}{1+\tau_\mathrm{d}/\tau_{\mathrm{L,f}}}
    \label{eqn:ud2}
\end{equation}

\begin{table}[H]
    \centering
    \begin{tabular}{|l|l|}
    \hline
    $\langle \mathrm{St} \rangle ^*$ & $\left[0.01, 0.1, 0.5, 1, 10, 50\right]$ \\
    \hline
    $\langle \mathrm{We} \rangle ^*$ & $\left[0.01, 4, 9\right]$ \\
    \hline
    $\mathrm{Oh}$ & $0.1$ \\
    \hline
    {Number of droplets} & {5000 for each case} \\
    \hline
    {Initial distribution} & {Uniform random} \\
    \hline
    \end{tabular}
    \caption{{Droplet parameters.}}
    \label{tab:drop}
\end{table}

\subsection{Effect of St and We on droplet dispersion}
\label{sec:StWe}
\begin{figure}[H]
    \centering
    \begin{subfigure}{0.32\linewidth}
        \centering
        \includegraphics[width=\linewidth]{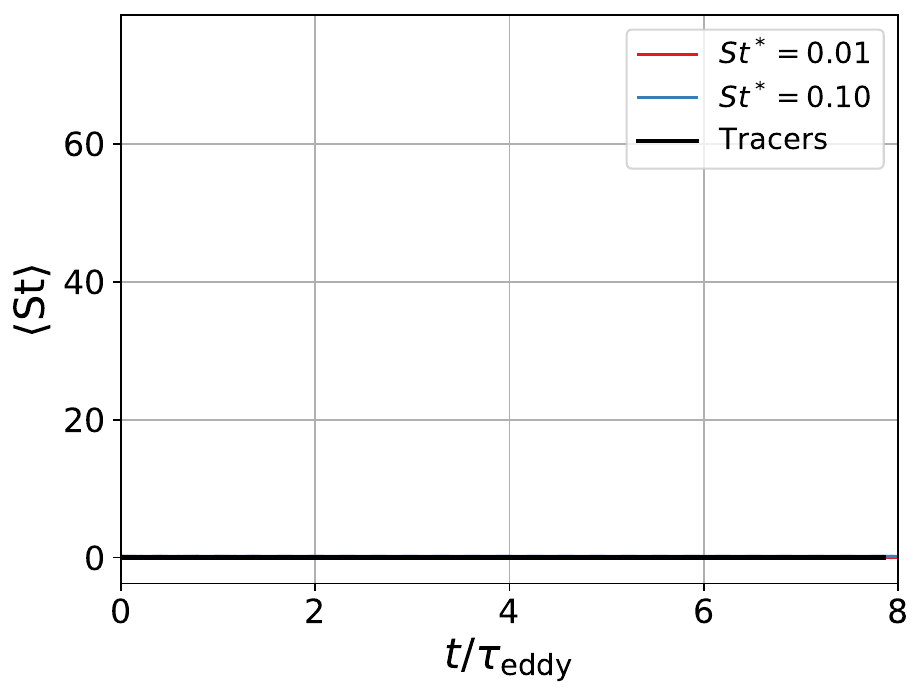}
        \caption{non-inertial}
        \label{fig:St_non_inertia}
    \end{subfigure}
    \hfill
    \begin{subfigure}{0.32\linewidth}
        \centering
        \includegraphics[width=\linewidth]{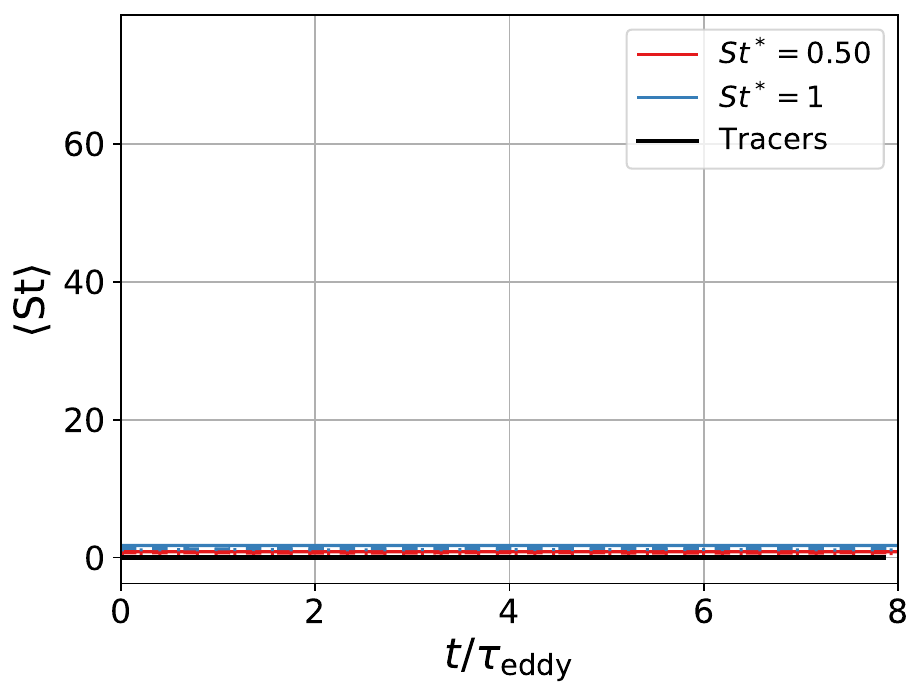}
        \caption{weakly-inertial}
        \label{fig:St_weak_inertia}
    \end{subfigure}
    \hfill
    \begin{subfigure}{0.32\linewidth}
        \centering
        \includegraphics[width=\linewidth]{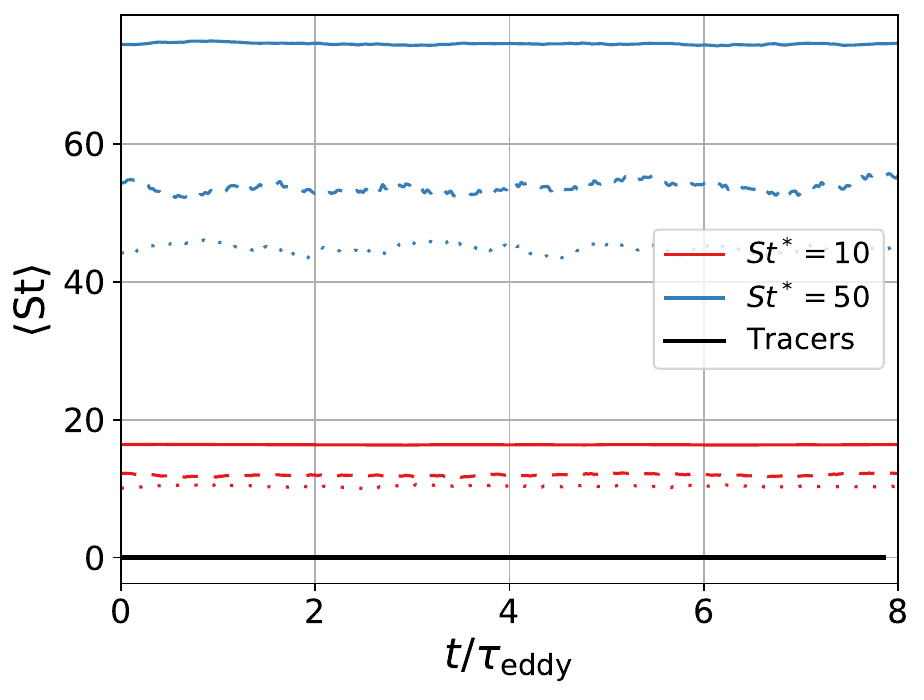}
        \caption{strongly-inertial}
        \label{fig:St_strong_inertia}
    \end{subfigure}

    \caption{{Mean Stokes number of droplets. Solid, dashed, and dotted lines correspond to $\mathrm{We}^*=0.01,\ 4,\ 9$, respectively. The line-styles and color codes of other figures in \Cref{sec:StWe} follow the conventions established in this figure.}}
    \label{fig:St}
\end{figure}

{Unlike the classically defined Stokes number based on Stokes drag law ($St^*$), the instantaneous Stokes number ($St$) is dynamically evaluated using the instantaneous droplet drag to capture the transient droplet deformation. Written explicitly, it satisfies,
\begin{align}
    St=\frac{\tau_d}{\tau_\eta}=\frac{\rho_\mathrm{d}}{\rho_\mathrm{f}}\frac{4}{3}\frac{d}{C_\mathrm{D}}\frac{1}{\left|\Delta\bm{u}\right|}\frac{1}{\tau_\eta}
\end{align}
}
As illustrated in \Cref{fig:St} and listed in \Cref{tab:StWe}, the average Stokes number, $\langle \mathrm{St} \rangle $, monotonically decreases with increasing $\langle \mathrm{We} \rangle $. While this reduction is insignificant for non-inertial droplets, it is highly pronounced for strongly-inertial droplets. This trend is driven by the increased projected frontal area and higher drag coefficient of the deformed droplets, which together reduce their relaxation times.

\begin{table}[H]
\centering
 \begin{tabular}{cccccccc}
    \hline
    $\langle\mathrm{St}\rangle^*$ & $\langle\mathrm{We}\rangle^*$ & $\langle \mathrm{St} \rangle$ & $\langle \mathrm{We} \rangle$ & $\langle \tau_\mathrm{L} \rangle/\tau_{\text{eddy}}$ & $\langle K \rangle$ & $\langle u_\mathrm{d}^2 \rangle$ & $\tau_\mathrm{decay}/\langle \tau_{\text{L}} \rangle$ \\
    \hline
    0.01   & 0.01  & 0.0178  & 0.00294  & 0.210 & 0.0174 & 0.478 & 0.0288 \\
    0.01   & 4.0  & 0.0163  & 0.782  & 0.228 & 0.0183 & 0.476 & 0.667 \\
    0.01   & 9.0  & 0.0153  & 1.27  & 0.238 & 0.0196 & 0.477 & 0.833 \\
    0.10   & 0.01  & 0.176  & 0.00701  & 0.197 & 0.0166 & 0.486 & 0.0313 \\
    0.10   & 4.0  & 0.154  & 1.66  & 0.253 & 0.0204 & 0.486 & 0.599 \\
    0.10   & 9.0  & 0.143  & 2.74  & 0.260 & 0.0222 & 0.483 & 0.738 \\
    0.50   & 0.01  & 0.843  & 0.0183  & 0.220 & 0.0179 & 0.475 & 0.0281 \\
    0.50   & 4.0  & 0.666  & 3.56  & 0.261 & 0.0216 & 0.484 & 0.569 \\
    0.50   & 9.0  & 0.601  & 5.78  & 0.254 & 0.0219 & 0.484 & 0.759 \\
    1.00   & 0.01  & 1.75  & 0.0298  & 0.277 & 0.0208 & 0.446 & 0.0224 \\
    1.00   & 4.0  & 1.32  & 4.49  & 0.269 & 0.0212 & 0.472 & 0.567 \\
    1.00   & 9.0  & 1.13  & 8.51  & 0.241 & 0.0191 & 0.474 & 0.873 \\
    10.00  & 0.01  & 16.4 & 0.0199  & 0.500 & 0.0214 & 0.260 & 0.0144 \\
    10.00  & 4.0  & 12.1 & 4.64  & 0.380 & 0.0199 & 0.326 &  0.460 \\
    10.00  & 9.0  & 10.4 & 9.05  & 0.354 & 0.0190 & 0.336 & 0.702 \\
    50.00  & 0.01  & 74.5 & 0.0164  & 1.14 & 0.0215 & 0.118 & 0.00924 \\
    50.00  & 4.0  & 54.2 & 5.00  & 0.676 & 0.0191 & 0.181 & 0.376 \\
    50.00  & 9.0  & 44.9 & 10.5 & 0.638 & 0.0195 & 0.197 & 0.565 \\
    \hline
  \end{tabular}
\caption{{Time-averaged statistics for $t/\tau_{\mathrm{eddy}} > 5$ (except $\langle \tau_\mathrm{L} \rangle/\tau_{\text{eddy}}$).}}
\label{tab:StWe}
\end{table}

\subsubsection{Statistics of Droplet Dispersion}
\label{sec:4.1.1}
\begin{figure}[H]
    \centering
    \begin{subfigure}{0.32\linewidth}
        \centering
        \includegraphics[width=\linewidth]{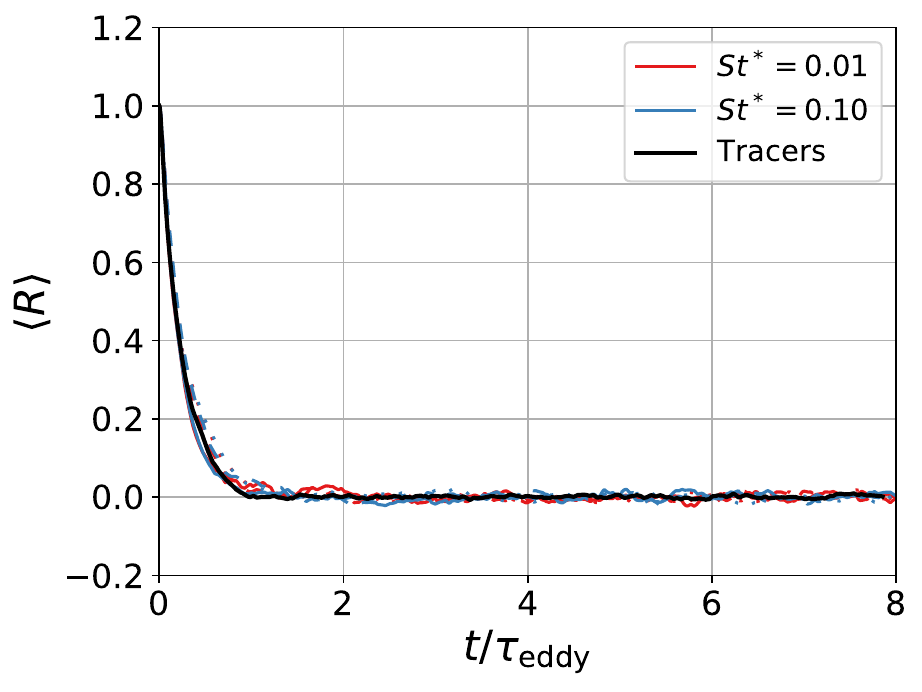}
        \caption{non-inertial}
        \label{fig:R_non_inertia_t}
    \end{subfigure}
    \hfill
    \begin{subfigure}{0.32\linewidth}
        \centering
        \includegraphics[width=\linewidth]{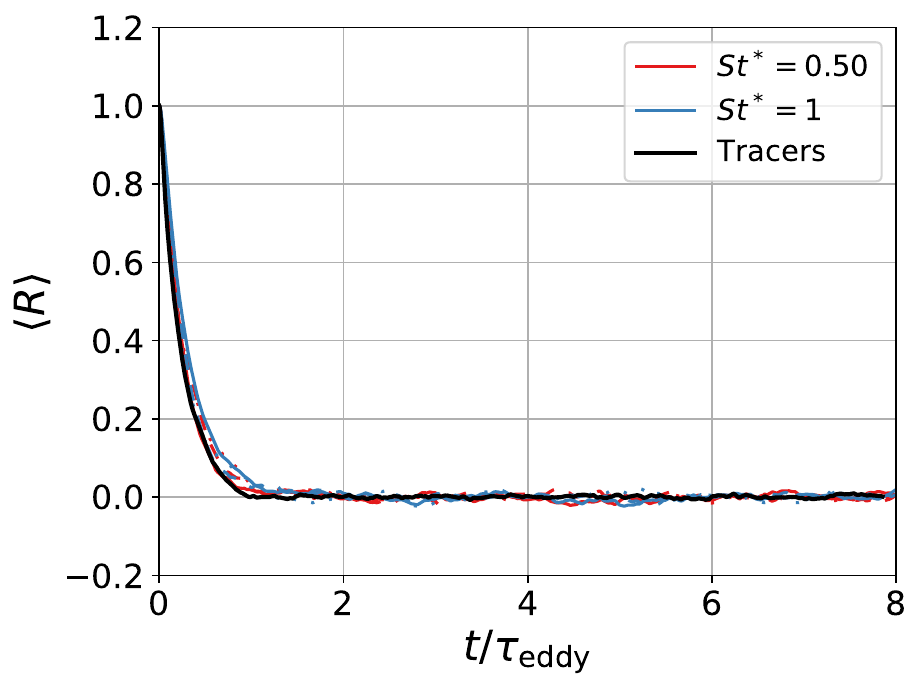}
        \caption{weakly-inertial}
        \label{fig:R_weak_inertia_t}
    \end{subfigure}
    \hfill
    \begin{subfigure}{0.32\linewidth}
        \centering
        \includegraphics[width=\linewidth]{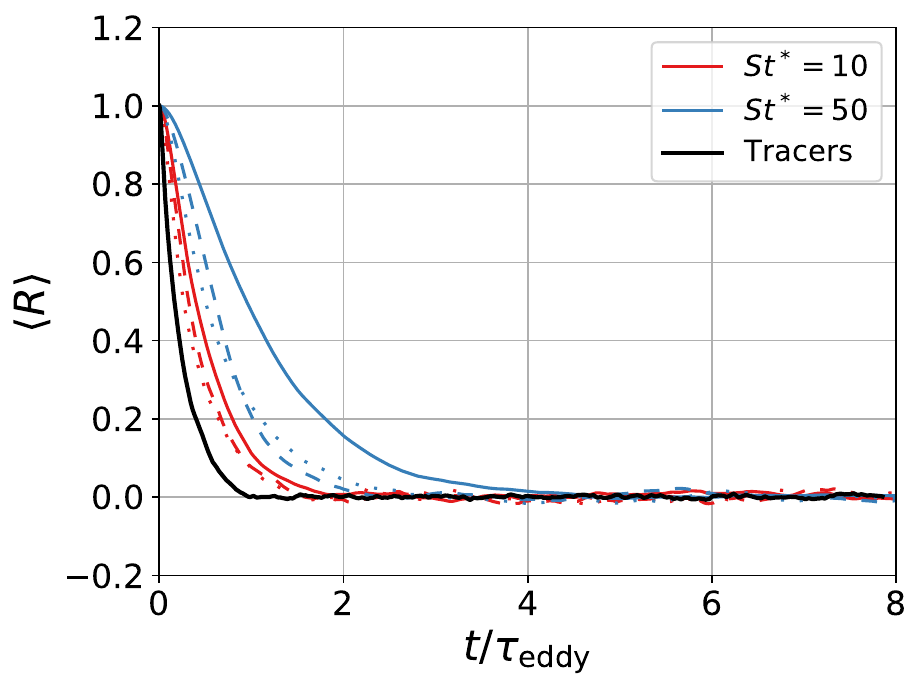}
        \caption{strongly-inertial}
        \label{fig:R_strong_inertia_t}
    \end{subfigure}
    \caption{Temporal autocorrelation of droplet velocity averaged over principal directions; time is normalized by the eddy turnover time $\tau_\mathrm{eddy}$.}
    \label{fig:R_t}
\end{figure}

\begin{figure}[H]
    \centering
    \begin{subfigure}{0.32\linewidth}
        \centering
        \includegraphics[width=\linewidth]{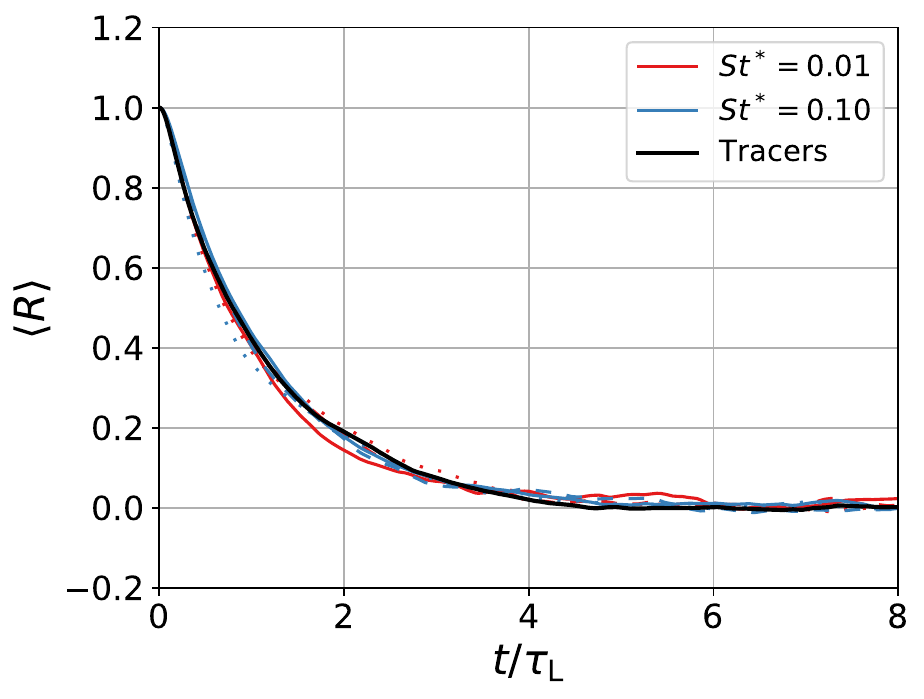}
        \caption{non-inertial}
        \label{fig:R_non_inertia_tau}
    \end{subfigure}
    \hfill
    \begin{subfigure}{0.32\linewidth}
        \centering
        \includegraphics[width=\linewidth]{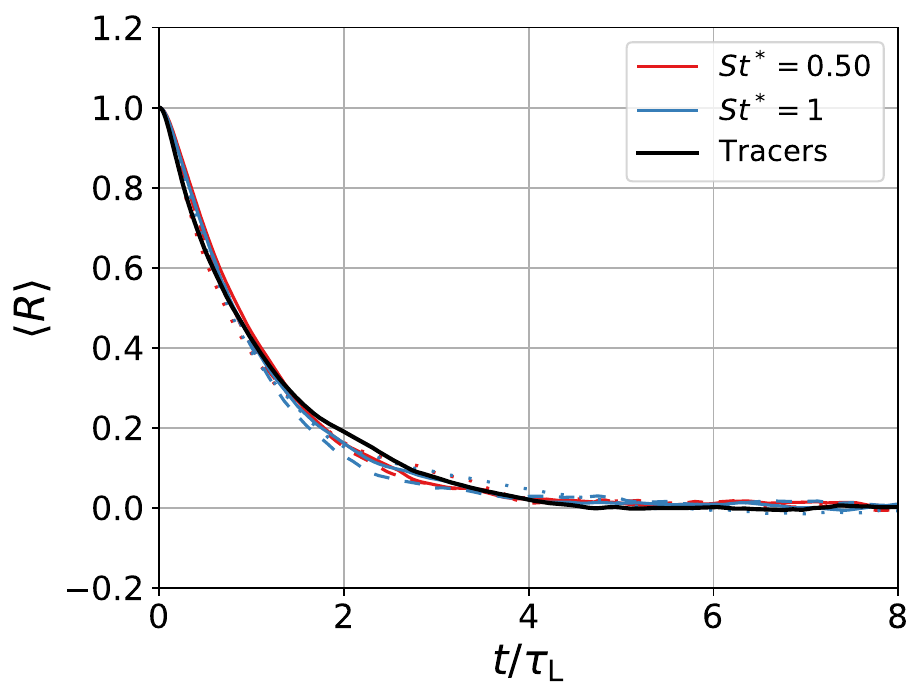}
        \caption{weakly-inertial}
        \label{fig:R_weak_inertia_tau}
    \end{subfigure}
    \hfill
    \begin{subfigure}{0.32\linewidth}
        \centering
        \includegraphics[width=\linewidth]{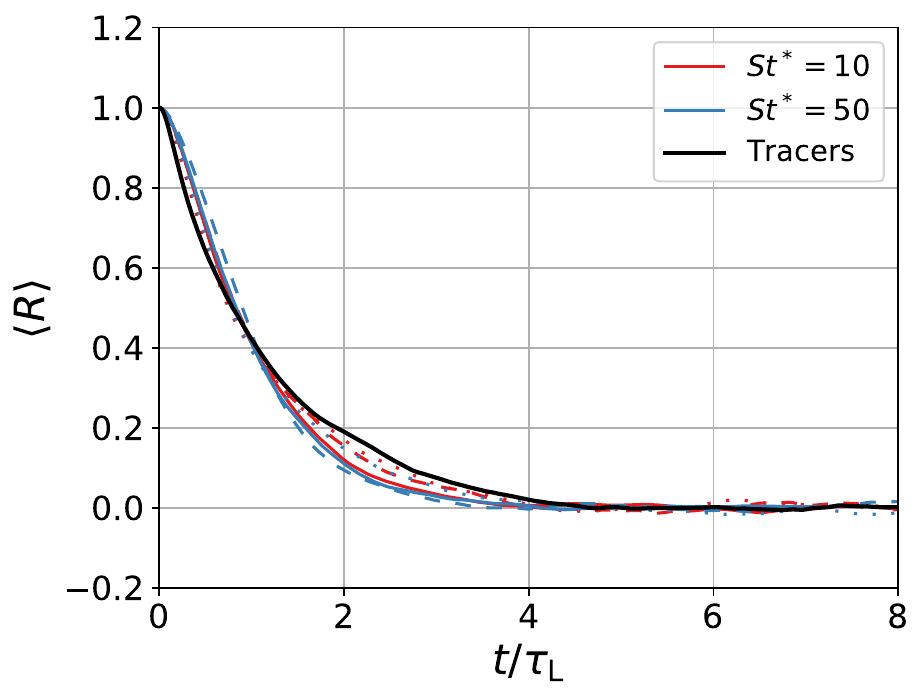}
        \caption{strongly-inertial}
        \label{fig:R_strong_inertia_tau}
    \end{subfigure}
    \caption{Temporal autocorrelation of droplet velocity averaged over principal directions; time is normalized by droplet Lagrangian time scale $\tau_\mathrm{L}$.}
    \label{fig:R_tau}
\end{figure}

The temporal autocorrelation of droplet velocity evaluated at different $\mathrm{St}^*$ and $\mathrm{We}^*$ is presented in \Cref{fig:R_t} and \Cref{fig:R_tau}. $R$ is defined as the average of diagonal elements of autocorrelation tensor $R_{ij}$. This metric quantifies the rate at which droplets lose memory of their initial velocities. As shown in \Cref{fig:R_t}, for non-inertial droplets, $R$ exhibits negligible differences across various $\mathrm{We}^*$, behaving closely to tracer particles. For weakly- and strongly-inertial droplets, droplets with higher $\mathrm{St}^*$ show a slower decay of $R$ compared to those with lower $\mathrm{St}^*$. Regarding the effect of deformation, strongly-inertial droplets with larger $\mathrm{We}^*$ show a faster decay of $R$ compared to those with smaller $\mathrm{We}^*$. This occurs because deformation increases aerodynamic drag, forcing the droplets to follow the local fluid motion more closely and consequently reducing the temporal velocity correlation. The effect of deformation is diminished for non- and weakly-inertial droplets, as they are less sensitive to changes of the drag force.

The Lagrangian time scale serves as a macroscopic measure of droplet velocity memory, quantifying the time required for its velocity to completely de-correlate from its initial state. From \Cref{tab:StWe} we can observe that $\tau_\mathrm{L}$ generally increases with increasing $\mathrm{St}^*$, which is consistent with the observation of $R$. The reason is that droplets with larger inertia are less responsive to high-frequency turbulent fluctuations \citep{squiresMeasurementsParticleDispersion1991}, resulting in a longer droplet Lagrangian time scale.

Since the velocity autocorrelation $R$ decays to 0 before $t/\tau_\mathrm{eddy}=5$ across all cases, long term statistics for all variables can be computed based on $t>5\tau_\mathrm{eddy}$. Regarding the effect of deformation, we note that for droplets with $\mathrm{St}^*<0.5$, $\tau_\mathrm{L}$ increases as $\langle \mathrm{We} \rangle $ increases, while for droplets with $\mathrm{St}^*>0.5$, $\tau_\mathrm{L}$ decreases as $\langle \mathrm{We} \rangle $ increases. At the transition point $\mathrm{St}^*=0.5$, it exhibits a non-monotonic dependence on $\langle \mathrm{We} \rangle $ where $\tau_\mathrm{L}/\tau_\mathrm{eddy}$ first increases to a peak value around 0.265 and then decreases. In addition, as shown in \Cref{fig:R_tau}, when time is normalized by $\tau_\mathrm{L}$, the curves of $R$ at different $\mathrm{St}^*$ and $\mathrm{We}^*$ collapse to a nearly universal curve with slight deviations. This result indicates that the Lagrangian time scale effectively captures the velocity memory of droplets across different cases, but it cannot fully account for the effect of droplet inertia and deformation, particularly for strongly inertial droplets.

\begin{figure}[H]
    \centering
    \begin{subfigure}{0.32\linewidth}
        \centering
        \includegraphics[width=\linewidth]{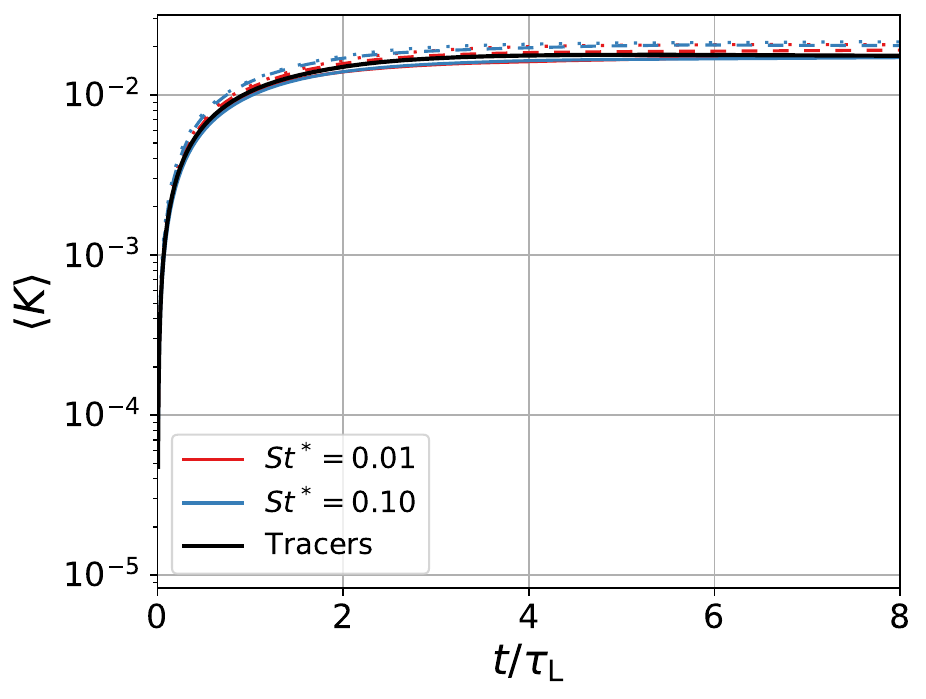}
        \caption{non-inertial}
        \label{fig:K_non_inertia}
    \end{subfigure}
    \hfill
    \begin{subfigure}{0.32\linewidth}
        \centering
        \includegraphics[width=\linewidth]{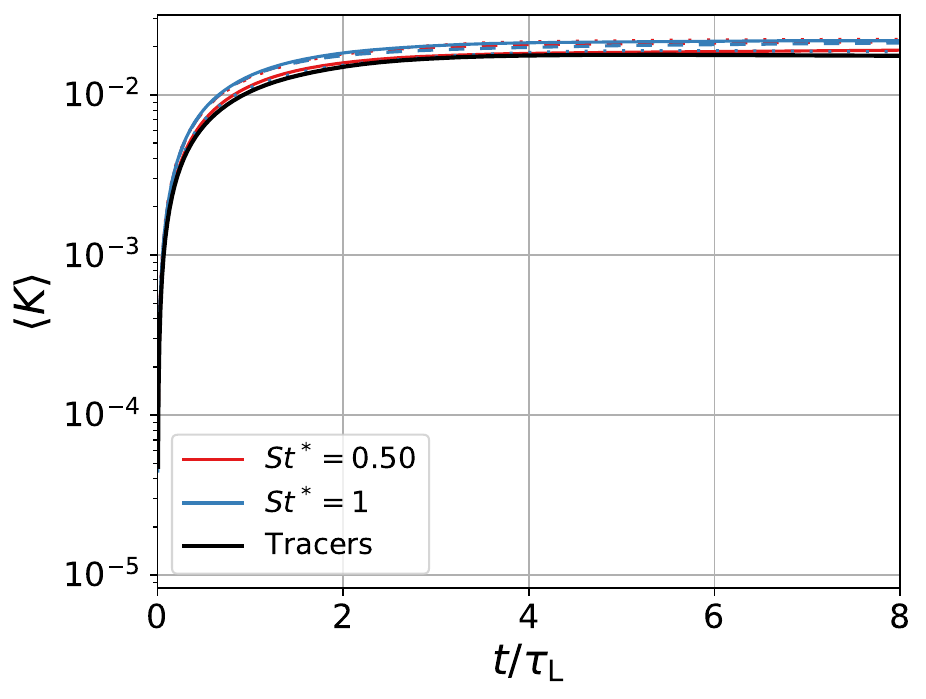}
        \caption{weakly-inertial}
        \label{fig:K_weak_inertia}
    \end{subfigure}
    \hfill
    \begin{subfigure}{0.32\linewidth}
        \centering
        \includegraphics[width=\linewidth]{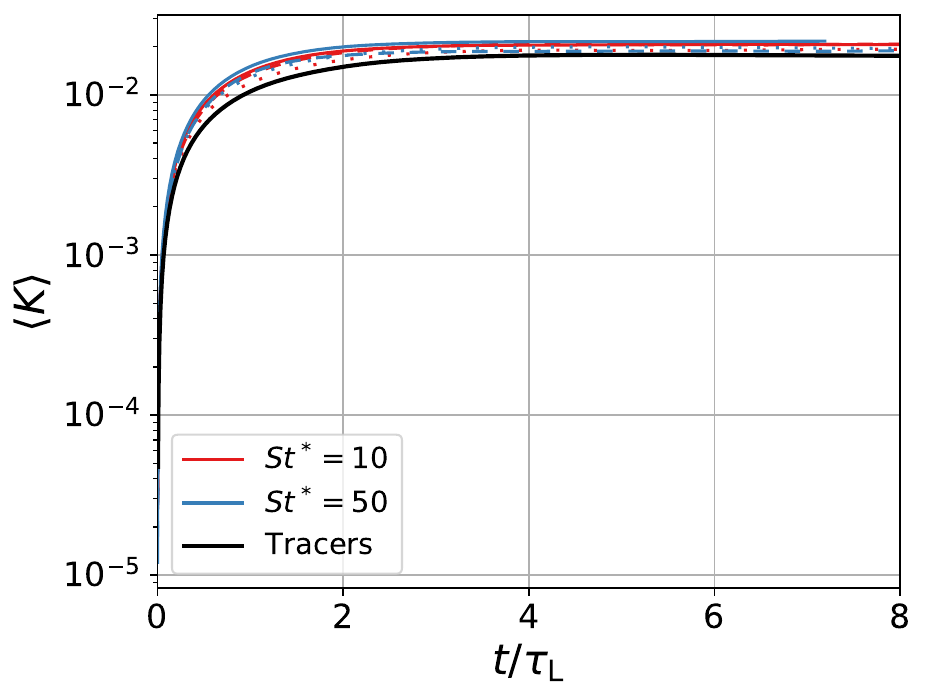}
        \caption{strongly-inertial}
        \label{fig:K_strong_inertia}
    \end{subfigure}
    \caption{Droplet dispersion coefficient.}
    \label{fig:K}
\end{figure}

{The droplet dispersion coefficient $K$ evaluated via \Cref{eqn:K} is presented in \Cref{fig:K}. The asymptotic values of $K$ at diffusive regime are summarized in \Cref{tab:StWe}. Consistent with the trends observed for $\langle \tau_\mathrm{L} \rangle $, $\langle K \rangle $ increases as $\langle \mathrm{We} \rangle $ increases for non-inertial droplets, but decreases for strongly-inertial droplets. The crossover in the Weber number dependency occurs in the weakly-inertial regime at $\langle \mathrm{St} \rangle ^*=1$. However, the deformation effect is less pronounced for strongly-inertial droplets due to a compensation mechanism between droplet velocity statistics and droplet Lagrangian time scale. According to \Cref{eqn:K_inf}, although $\tau_\mathrm{L}$ is reduced significantly due to deformation, the increased drag simultaneously increases mean-square droplet velocity as shown in \Cref{tab:StWe}, and therefore the product of these two competing factors ultimately results in a relatively weak sensitivity of $K$ to droplet deformation.}

To quantify the macroscopic transport of droplets relative to the turbulent flow, we evaluate reduced dispersion coefficient defined in \Cref{eqn:S} and \Cref{eqn:S_inf}, following the theoretical framework developed by \citet{gouesbetDispersionDiscreteParticles1984}. The results of $\langle S_\infty \rangle $ suggest that for droplets with $\mathrm{St}^*<1$, deformation will monotonically enhance droplet dispersion to make them disperse more rapidly than fluid particles. For $\mathrm{St}^*\ge1$, deformation will suppress droplet dispersion, but the dependence on Weber number is non-monotonic. Since results of $\langle S \rangle $ have the same trend as $\langle S_\infty \rangle $, it is omitted for brevity.

\begin{figure}[H]
    \centering
    \begin{subfigure}{0.32\linewidth}
        \centering
        \includegraphics[width=\linewidth]{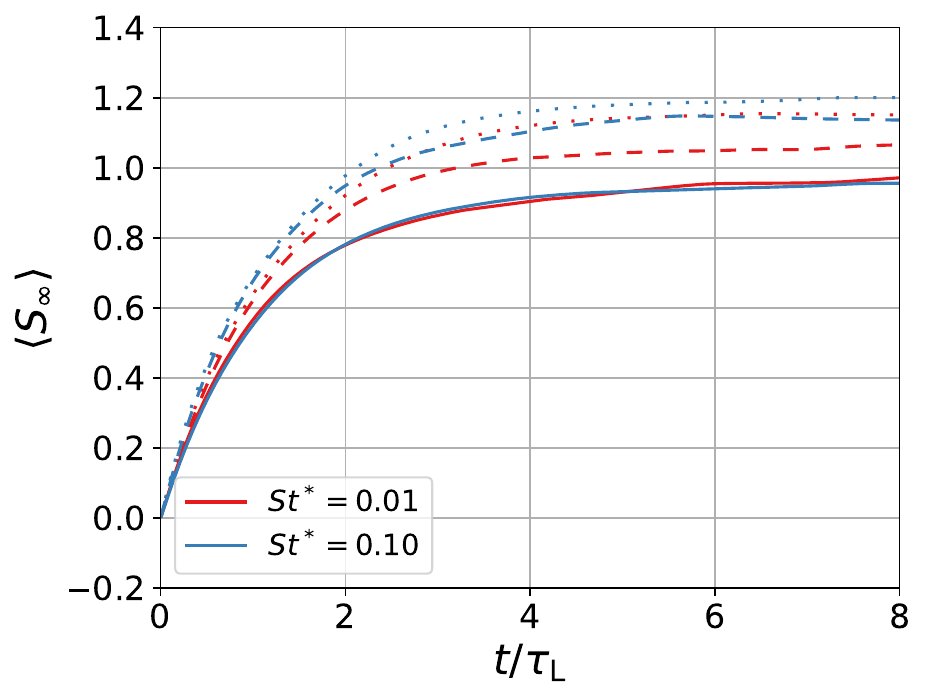}
        \caption{non-inertial}
        \label{fig:S_inf_non_inertia}
    \end{subfigure}
    \hfill
    \begin{subfigure}{0.32\linewidth}
        \centering
        \includegraphics[width=\linewidth]{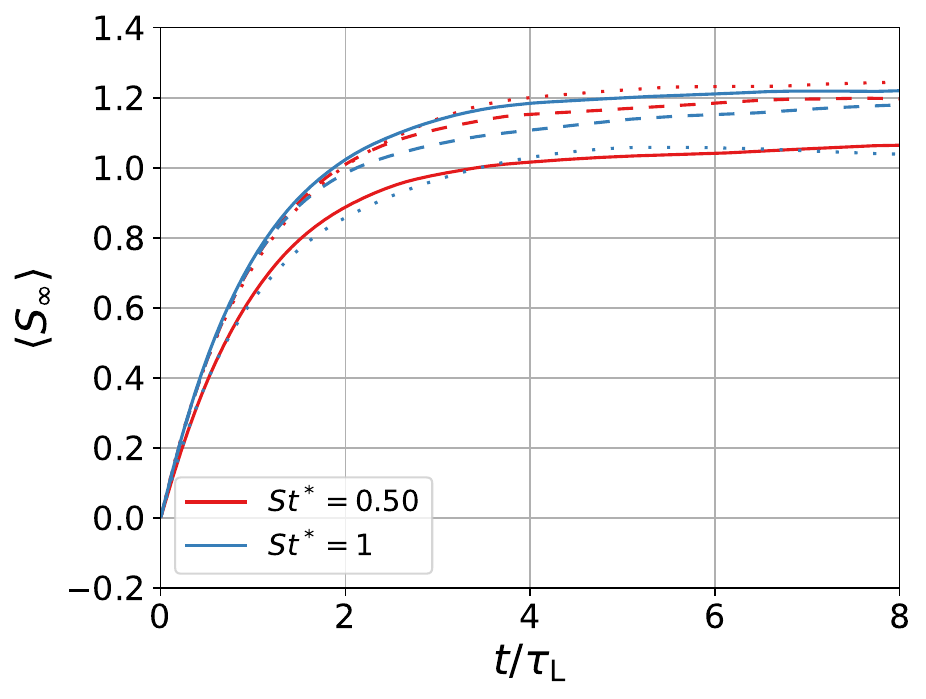}
        \caption{weakly-inertial}
        \label{fig:S_inf_weak_inertia}
    \end{subfigure}
    \hfill
    \begin{subfigure}{0.32\linewidth}
        \centering
        \includegraphics[width=\linewidth]{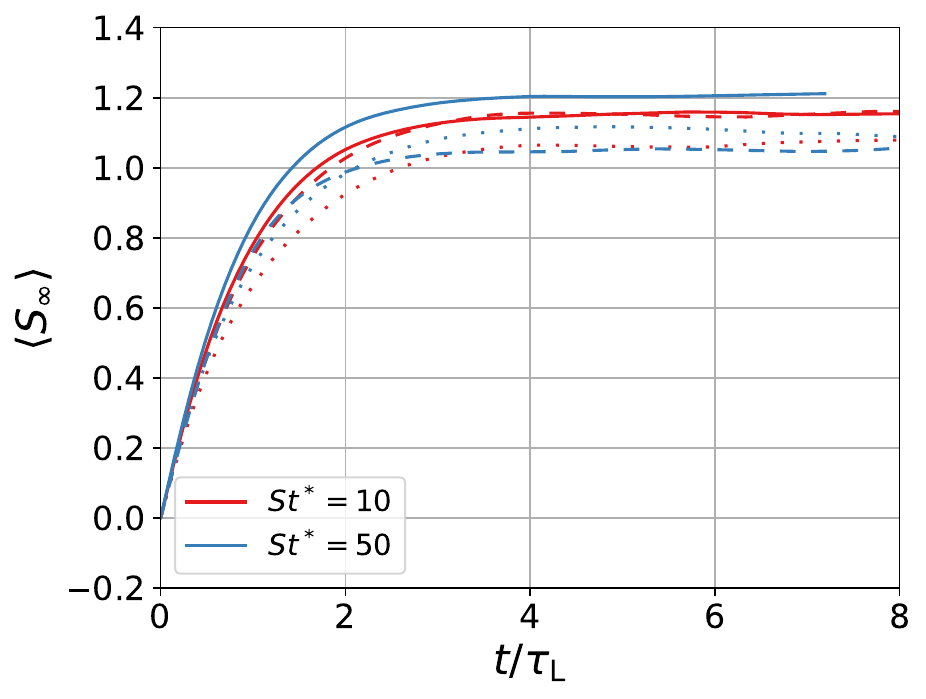}
        \caption{strongly-inertial}
        \label{fig:S_inf_strong_inertia}
    \end{subfigure}
    \caption{Reduced dispersion coefficient $S_\infty$.}
    \label{fig:S_inf}
\end{figure}

It is noteworthy that for non-inertial droplets, the dispersion is enhanced by deformation even with a decreased Stokes number, as shown in \Cref{tab:StWe} and \Cref{fig:K_non_inertia}. This result appears to contradict the classical linear theory \citep{wangDispersionHeavyParticles1993}, where a smaller Stokes number decreases the Lagrangian time scale for $\mathrm{St\rightarrow0}$. To understand this, we examine the unsteady deformation dynamics.
The TAB model describes a damped linear oscillator with a damping ratio $\zeta$
\begin{equation}
    \zeta = \frac{5}{4}\mathrm{Oh},
    \label{eqn:zeta}
\end{equation}
which equals 0.125 across all cases. The characteristic decay time $\tau_\mathrm{decay}$ of the shape oscillation is therefore $\tau_\mathrm{decay} = \tau_\mathrm{n}/\zeta$, where $\tau_\mathrm{n}$ is the undamped natural period associated to \Cref{TAB}. The ratio $\tau_\mathrm{decay}/\langle\tau_\mathrm{L}\rangle$ is listed in the last column of \Cref{tab:StWe}. For a fixed $\mathrm{St}^*$, $\tau_\mathrm{decay}/\langle\tau_\mathrm{L}\rangle$ increases as the Weber number increases, and the value lies between 0.3 to 0.9 for deformed droplets. The results indicate that the oscillation decay time for deformed droplets is comparable to the droplet Lagrangian time scale. For non-inertial droplets, the damped shape oscillation acts as a persistent perturbation that prolongs the velocity decorrelation process, leading to an increased Lagrangian time scale. This effect is negligible for strongly-inertial droplets, because the decorrelation is dominated by the drastic reduction of the droplet relaxation time. However, the re-introduction of inertia due to surface tension could potentially explain some of the non-monotonic Weber number trends previously observed and it will be further investigated in \Cref{sec:TABunsteady}.


\subsection{Clustering and droplet-vortex interaction}
The macroscopic dispersion statistics reported in \Cref{sec:StWe} alone do not describe the uniformity of spatial distribution of droplets. As introduced in \Cref{sec:clustering}, inertial droplets tend to cluster in regions of low vorticity and high strain rate, which is known as preferential concentration. In this section, we use the method introduced in \Cref{sec:sub:moffatt} as well as Voronoi analysis to investigate the mechanisms of clustering and how it is modulated by droplet deformation.

\subsubsection{Statistics of Direction Cosines in Local Vortex Coordinate}
\label{sec:cos}
\begin{figure}[H]
    \centering
    \begin{subfigure}{0.32\linewidth}
        \centering
        \includegraphics[width=\linewidth]{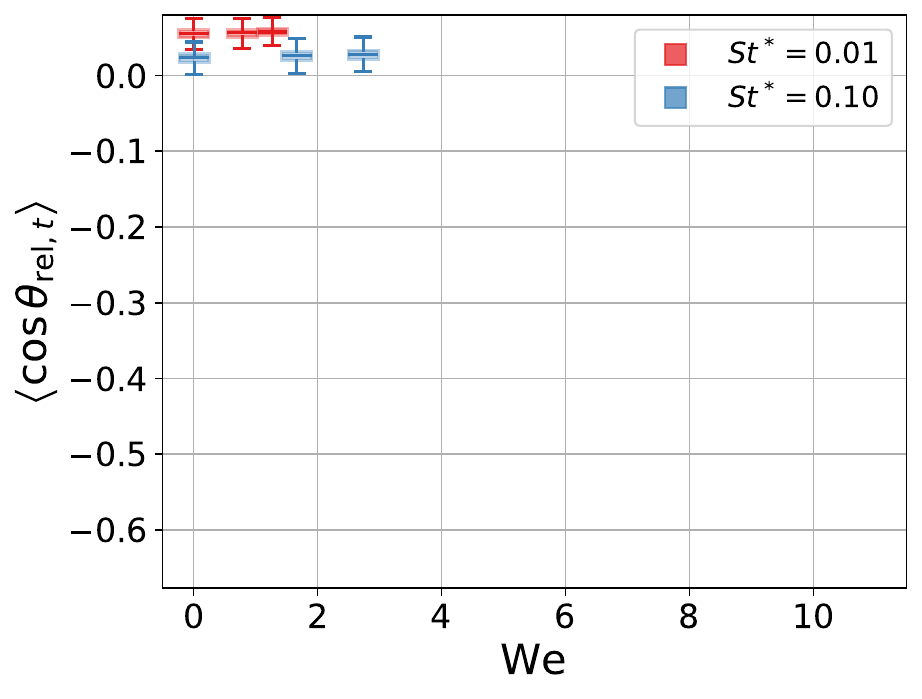}
        \caption{non-inertial}
        \label{fig:cos_t_non_inertia}
    \end{subfigure}
    \hfill
    \begin{subfigure}{0.32\linewidth}
        \centering
        \includegraphics[width=\linewidth]{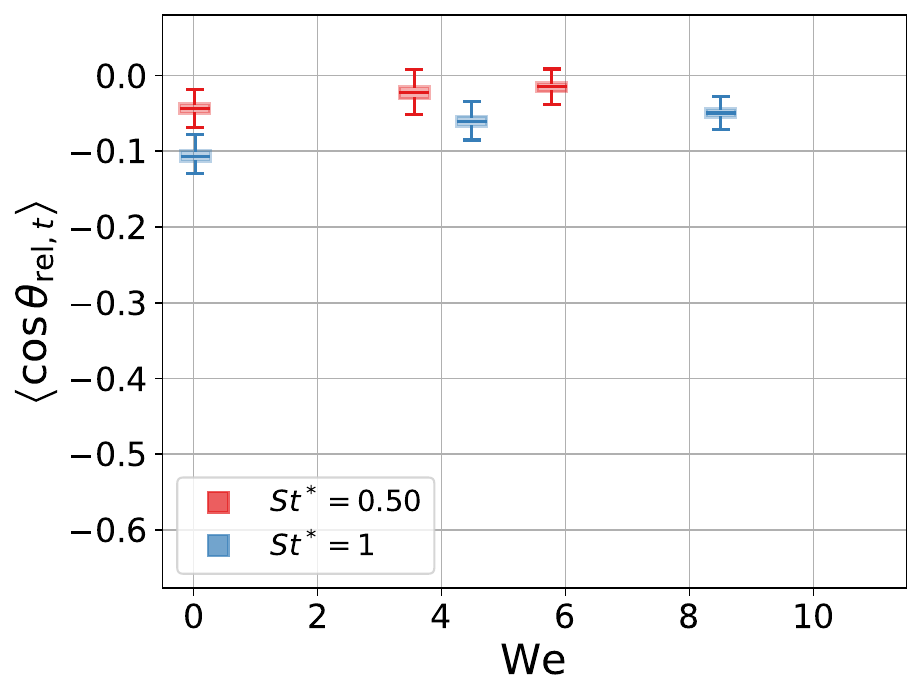}
        \caption{weakly-inertial}
        \label{fig:cos_t_weak_inertia}
    \end{subfigure}
    \hfill
    \begin{subfigure}{0.32\linewidth}
        \centering
        \includegraphics[width=\linewidth]{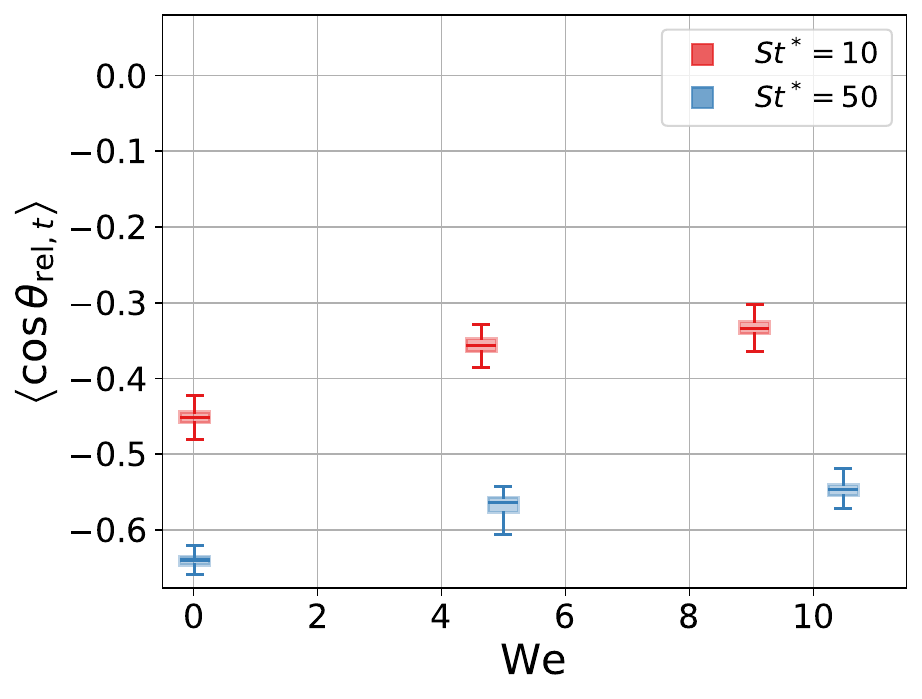}
        \caption{strongly-inertial}
        \label{fig:cos_t_strong_inertia}
    \end{subfigure}
    \caption{Time-averaged relative droplet velocity component tangential to the local fluid flow ($t$-direction).}
    \label{fig:cos_t}
\end{figure}

Direction cosines of droplet relative velocity in the local vortex coordinate system provide deeper insights into droplet dispersion mechanisms. The direction cosines in $t$-direction are plotted in {\Cref{fig:cos_t}}. The statistics are presented using whisker-box plots, where the central box represents the inter-quartile range (IQR) and the whiskers extend from the box to the furthest data point lying within 1.5 times the IQR from the box. The results show that for non-inertial and weakly-inertial droplets $\langle \cos\theta_{\mathrm{rel},t} \rangle $ is close to 0, indicating that these droplets are following the swirling motion of eddies closely. For strongly-inertial droplets, $\langle \cos\theta_{\mathrm{rel},t} \rangle $ becomes substantially negative, in contrast to the near-zero values for lower-inertia droplets. This negative value signifies that strongly-inertial droplets resist the swirling motion of eddies and tend to lag behind the fluid. In addition, we note that increasing Weber number makes $\langle \cos\theta_{\mathrm{rel},t} \rangle $ less negative for strongly-inertial droplets. The enhanced drag associated with deformation allows these droplets to respond more readily to the swirling flow, driving $\langle \cos\theta_{\mathrm{rel},t} \rangle $ toward zero. In addition, the small range of IQR and short whiskers indicate that the statistics of $\langle \cos\theta_{\mathrm{rel},t} \rangle $ is very stable. 

\begin{figure}[H]
    \centering
    \begin{subfigure}{0.32\linewidth}
        \centering
        \includegraphics[width=\linewidth]{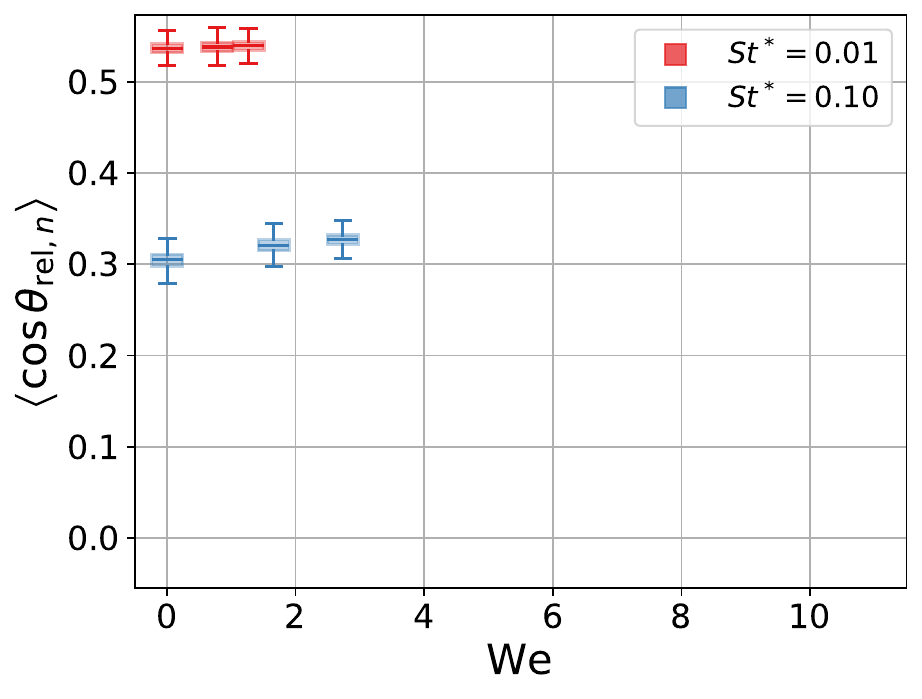}
        \caption{non-inertial}
        \label{fig:cos_n_non_inertia}
    \end{subfigure}
    \hfill
    \begin{subfigure}{0.32\linewidth}
        \centering
        \includegraphics[width=\linewidth]{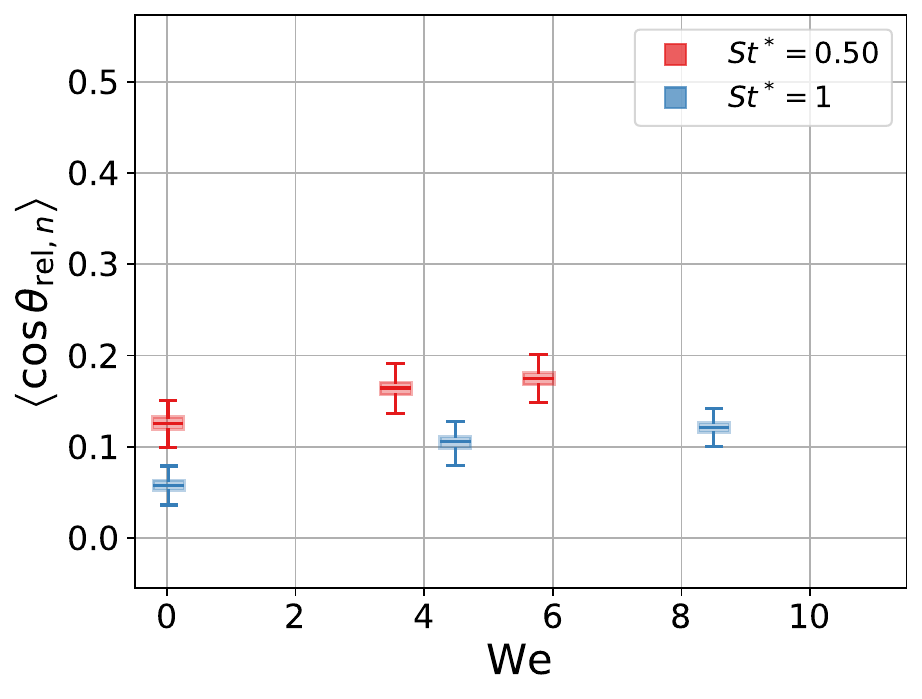}
        \caption{weakly-inertial}
        \label{fig:cos_n_weak_inertia}
    \end{subfigure}
    \hfill
    \begin{subfigure}{0.32\linewidth}
        \centering
        \includegraphics[width=\linewidth]{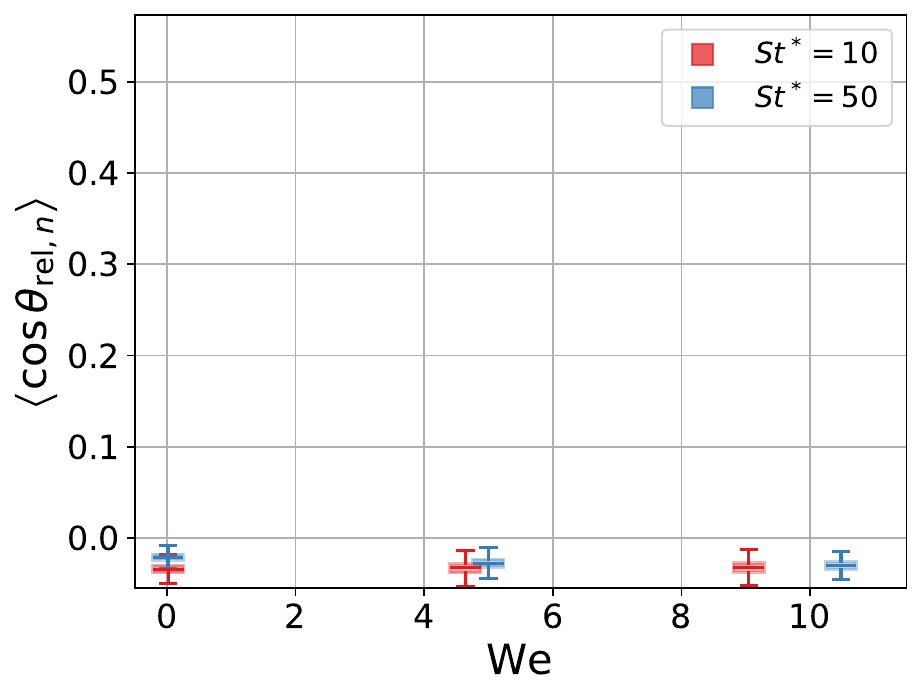}
        \caption{strongly-inertial}
        \label{fig:cos_n_strong_inertia}
    \end{subfigure}
    \caption{Time-averaged relative droplet velocity component normal to the other two components ($n$-direction).}
    \label{fig:cos_n}
\end{figure}

The normal direction cosine $\langle \cos\theta_{\mathrm{rel},n} \rangle $ shown in \Cref{fig:cos_n} decreases monotonically with Stokes number. Both non-inertial and weakly-inertial droplet demonstrate an outward centrifuging response to the vortex, as expected. The peak relative centrifuging velocity appears to occur for the non-inertial droplets. However it is important to note that this quantity is normalized in terms of vanishingly small relative velocity. Accordingly, the absolute value of this preferential motion is small for these droplets.

For strongly-inertial droplets, $\langle \cos\theta_{\mathrm{rel},n} \rangle $ becomes a small negative value but is statistically distinguishable from 0, indicating that centrifugal effect is significantly suppressed, and strongly-inertial droplets slightly lag behind the outward centrifuging fluids. The effect of deformation on $\langle \cos\theta_{\mathrm{rel},n} \rangle $ is almost negligible, although we note a slight enhancement on $\langle \cos\theta_{\mathrm{rel},n} \rangle $ for non-inertial and weakly-inertial droplets, and slight suppression for strongly-inertial droplets. The statistics of $\langle \cos\theta_{\mathrm{rel},n} \rangle $ is very stable as well.

\begin{figure}[H]
    \centering
    \begin{subfigure}{0.32\linewidth}
        \centering
        \includegraphics[width=\linewidth]{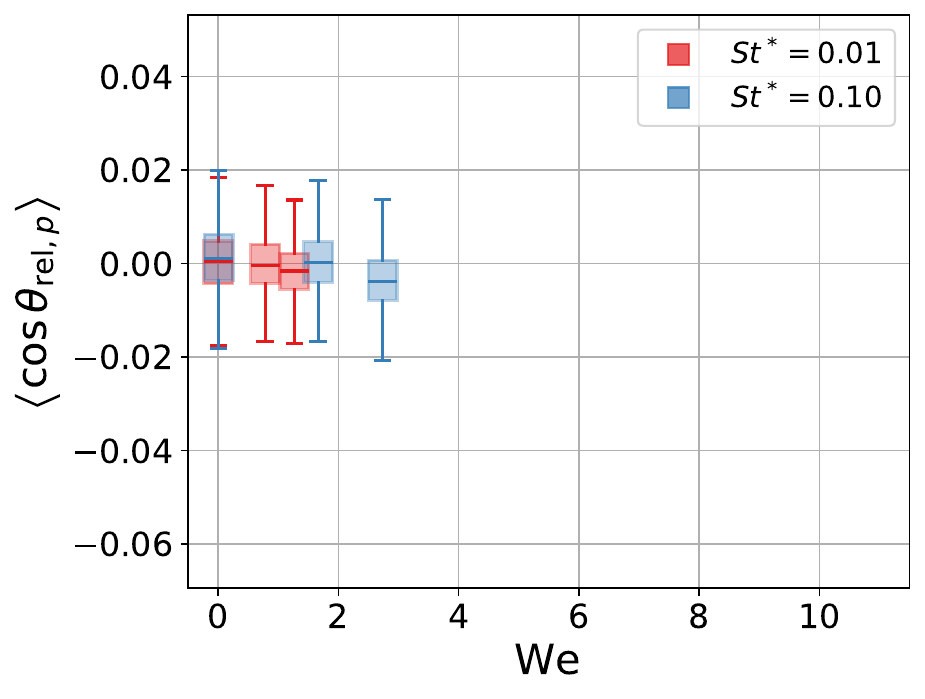}
        \caption{non-inertial}
        \label{fig:cos_p_non_inertia}
    \end{subfigure}
    \hfill
    \begin{subfigure}{0.32\linewidth}
        \centering
        \includegraphics[width=\linewidth]{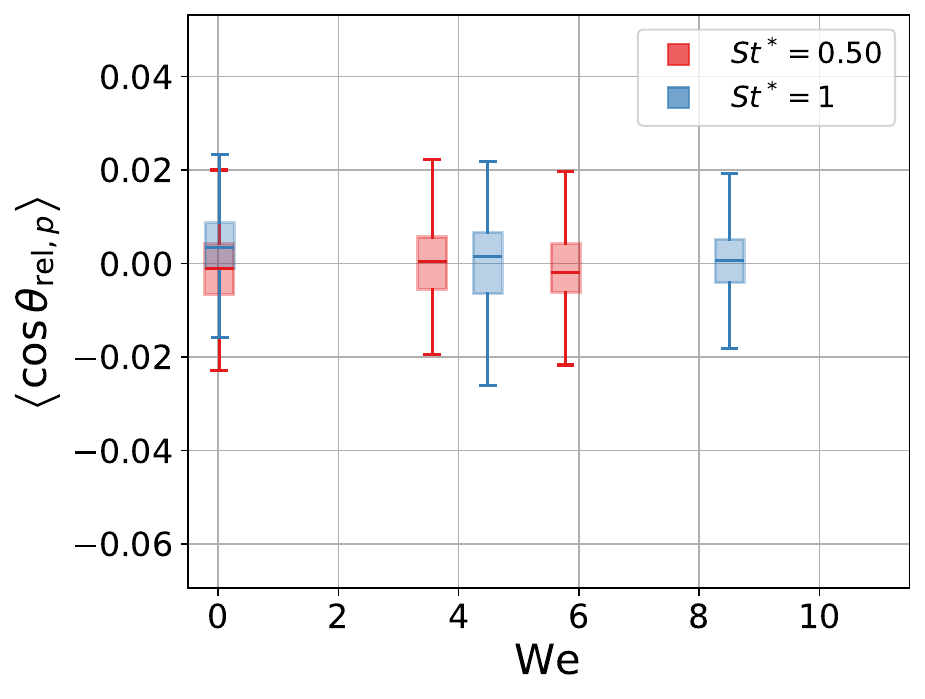}
        \caption{weakly-inertial}
        \label{fig:cos_p_weak_inertia}
    \end{subfigure}
    \hfill
    \begin{subfigure}{0.32\linewidth}
        \centering
        \includegraphics[width=\linewidth]{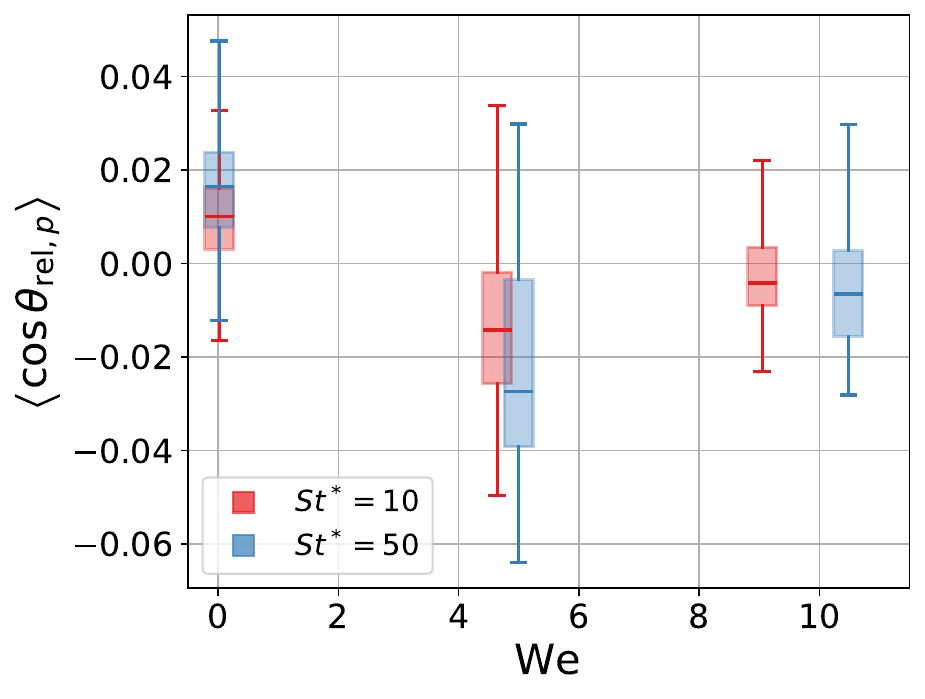}
        \caption{strongly-inertial}
        \label{fig:cos_p_strong_inertia}
    \end{subfigure}
    \caption{Time-averaged relative droplet velocity component parallel to the local vorticity vector ($p$-direction).}
    \label{fig:cos_p}
\end{figure}

Along the $p$-direction shown in \Cref{fig:cos_p}, the distributions of $\langle \cos\theta_{\mathrm{rel},p}\rangle$ for the non-inertial and weakly-inertial droplets remain centered near zero across all Weber number cases. However, their whiskers are noticeably longer than those of $\langle \cos\theta_{\mathrm{rel},t}\rangle$ and $\langle \cos\theta_{\mathrm{rel},n}\rangle$, indicating stronger fluctuations of the relative motion along the vortex axis. For strongly-inertial droplets, the distribution of $\langle \cos\theta_{\mathrm{rel},p}\rangle$ deviates slightly from zero and exhibits even longer whiskers than the lower-inertia cases, suggesting enhanced axial dispersion. The near-zero central tendency indicates that these axial motions do not correspond to a persistent directional drift.

\subsubsection{{Voronoi Analysis}}
\begin{figure}[H]
    \centering
    \begin{subfigure}{0.32\linewidth}
        \centering
        \includegraphics[width=\linewidth]{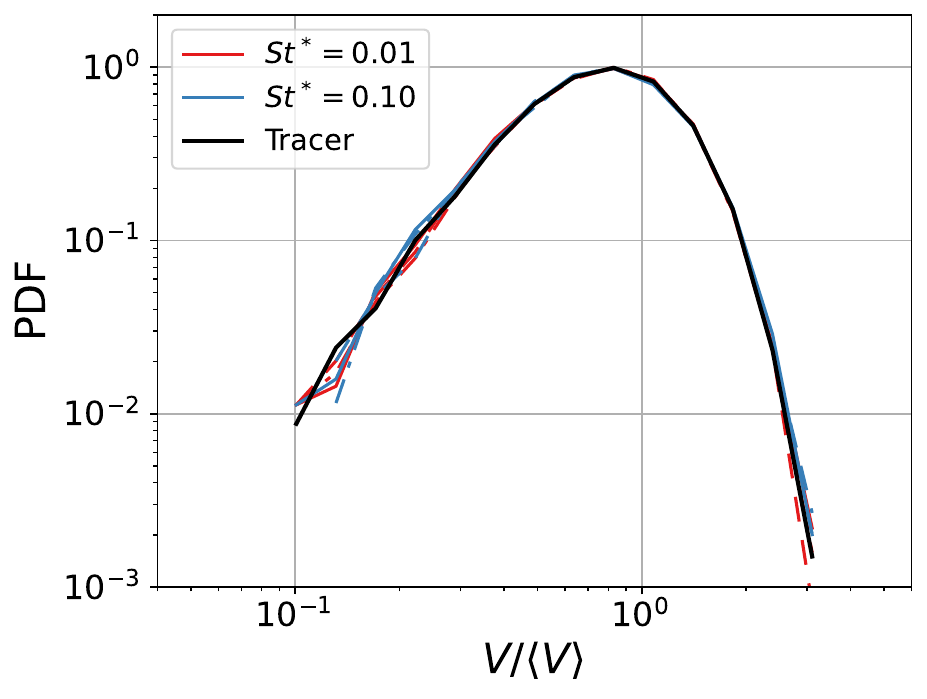}
        \caption{non-inertial}
        \label{fig:pdf_non_inertia}
    \end{subfigure}
    \hfill
    \begin{subfigure}{0.32\linewidth}
        \centering
        \includegraphics[width=\linewidth]{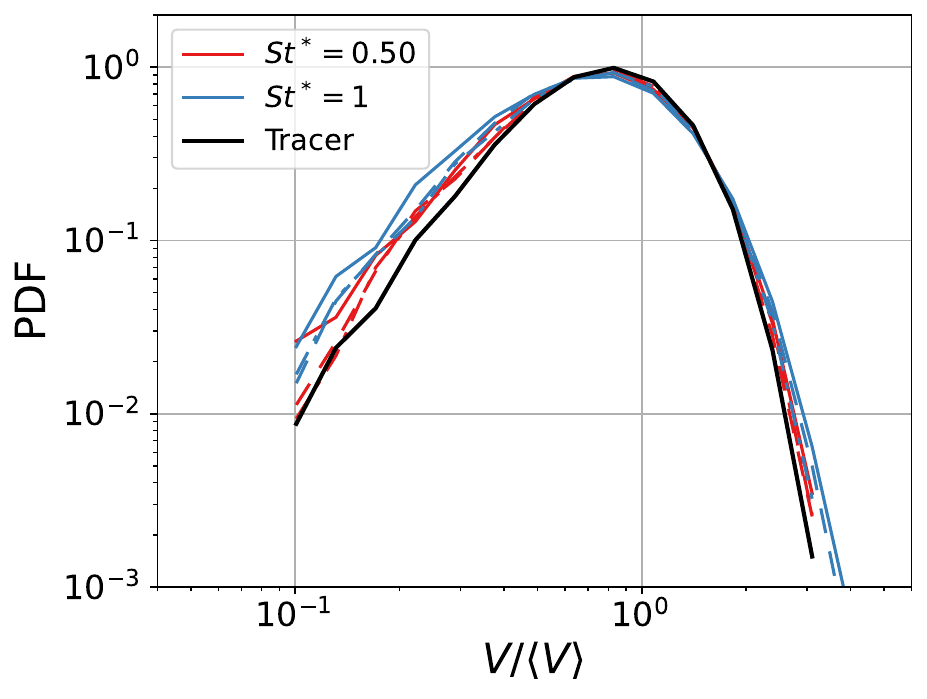}
        \caption{weakly-inertial}
        \label{fig:pdf_weak_inertia}
    \end{subfigure}
    \hfill
    \begin{subfigure}{0.32\linewidth}
        \centering
        \includegraphics[width=\linewidth]{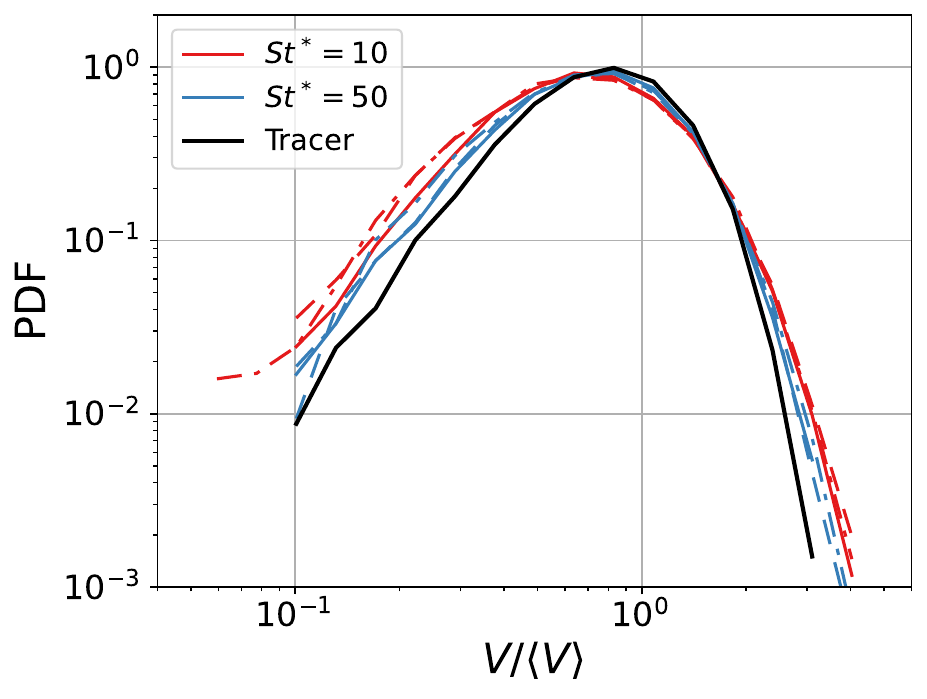}
        \caption{strongly-inertial}
        \label{fig:pdf_strong_inertia}
    \end{subfigure}
    \caption{Probability density function of Voronoi cell volume.}
    \label{fig:voronoi_pdf}
\end{figure}

To examine the strength of clustering, the probability density function (PDF) of the normalized Voronoi cell volume ${V}/{\langle V \rangle }$ is shown in \Cref{fig:voronoi_pdf}. For non-inertial droplets, the clustering is weak, as their PDFs are nearly identical to those of the fluid particles. For weak- and strongly-inertial droplets, preferential concentration becomes evident, as the PDFs exceed those of the fluid particle at small ${V}/{\langle V \rangle }$. Deformation is observed to weaken preferential concentration for weakly-inertial droplets, while enhancing it for strongly-inertial droplets. This indicates that deformation effect on clustering is dependent on droplet inertia. It suppresses clustering in the weakly-inertial regime but enhances it in strongly-inertial regime.

\begin{figure}[H]
    \centering
    \includegraphics[width=0.5\linewidth]{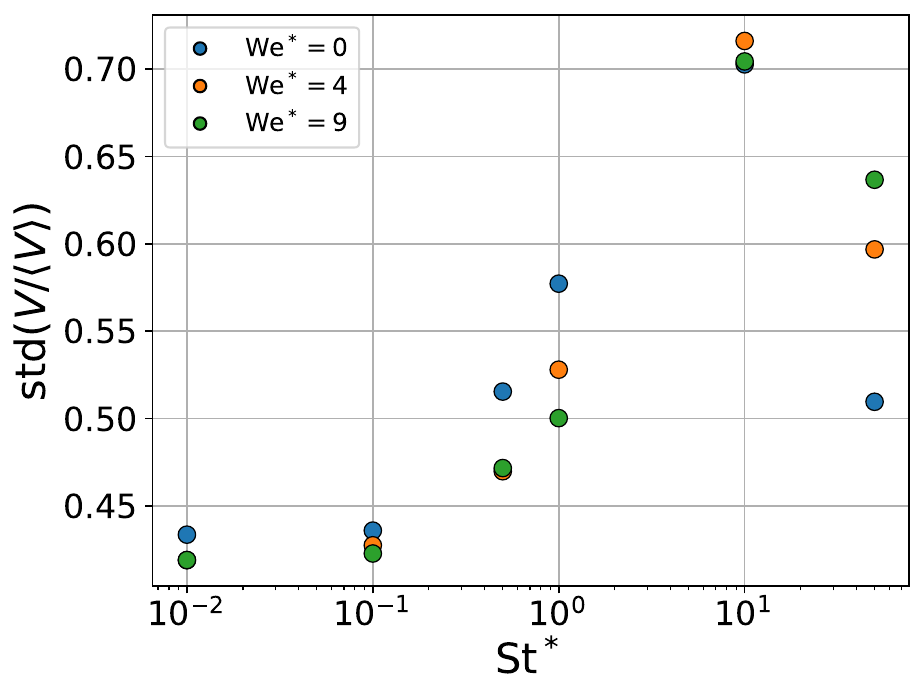}
    \caption{Standard deviation of normalized Voronoi cell volume.}
    \label{fig:std_V}
\end{figure}

\begin{figure}[H]
    \centering
    \begin{subfigure}{0.32\linewidth}
        \centering
        \includegraphics[width=\linewidth]{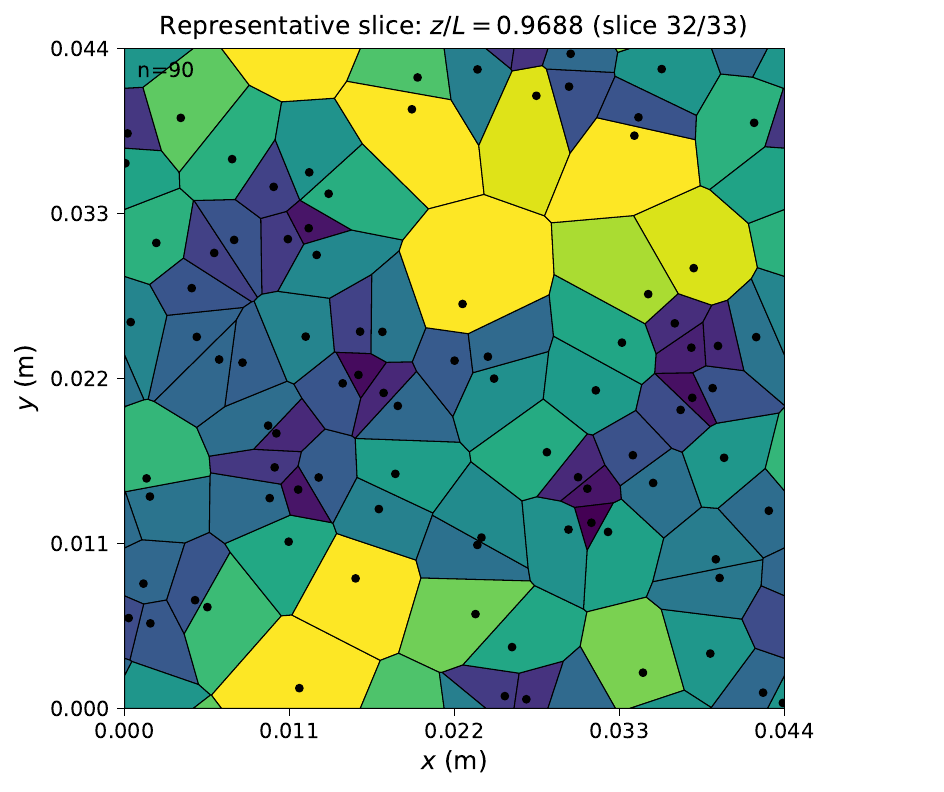}
        \caption{$\mathrm{We}^*=0.01$}
        \label{fig:voronoi_St50We00}
    \end{subfigure}
    \hfill
    \begin{subfigure}{0.32\linewidth}
        \centering
        \includegraphics[width=\linewidth]{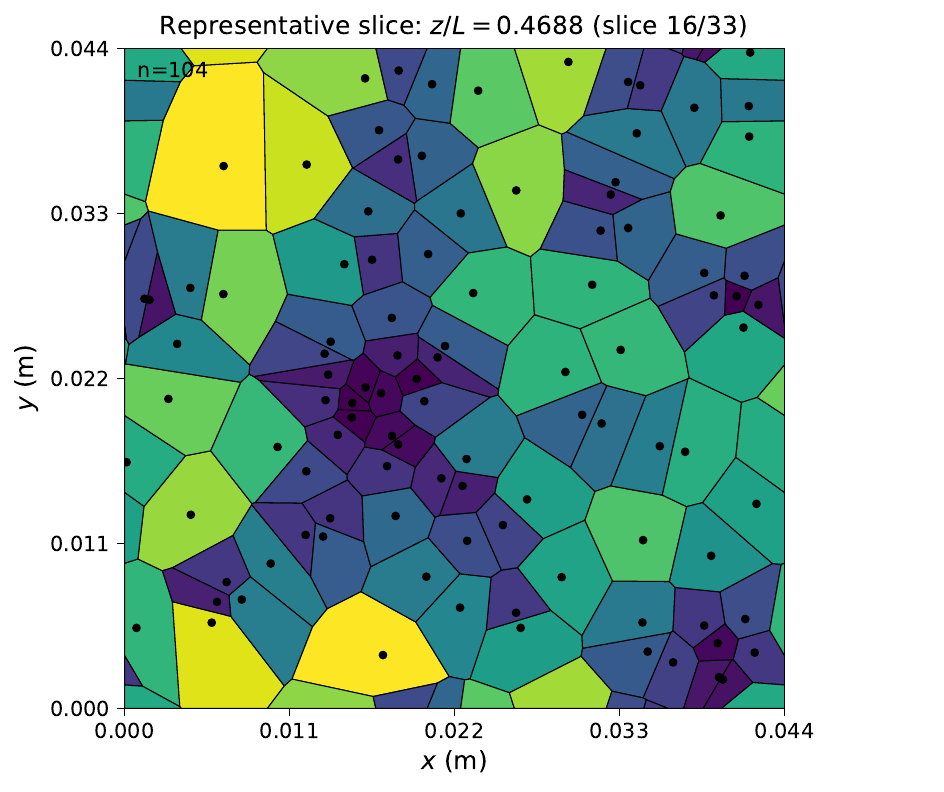}
        \caption{$\mathrm{We}^*=4$}
        \label{fig:voronoi_St50We04}
    \end{subfigure}
    \hfill
    \begin{subfigure}{0.32\linewidth}
        \centering
        \includegraphics[width=\linewidth]{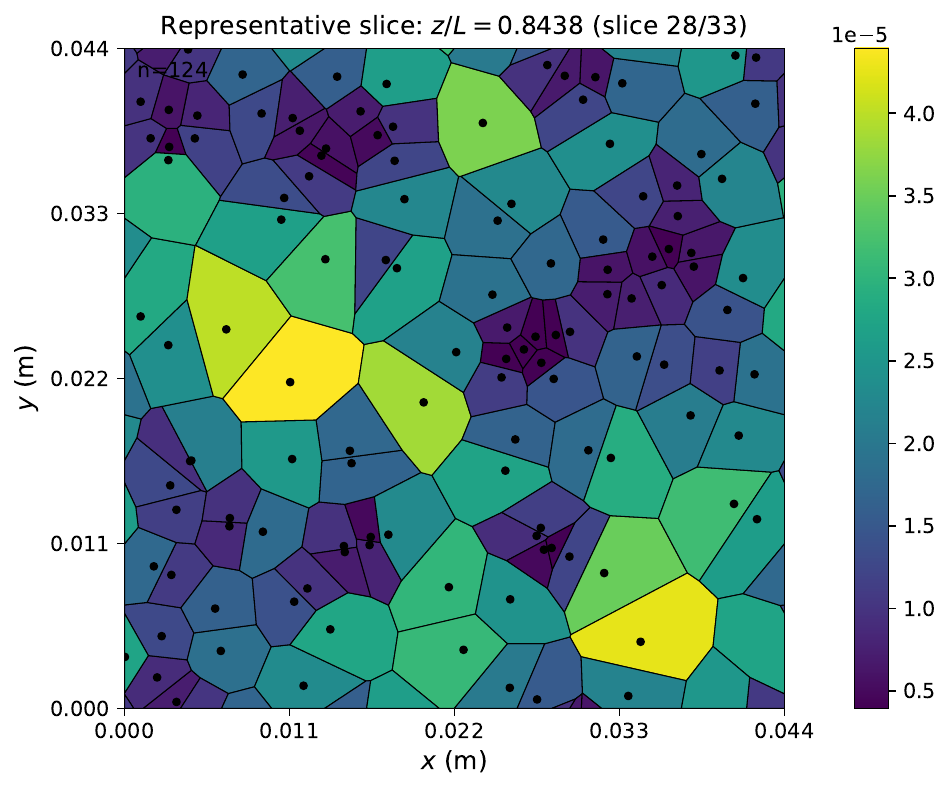}
        \caption{$\mathrm{We}^*=9$}
        \label{fig:voronoi_St50We09}
    \end{subfigure}
    \caption{Two-dimensional Voronoi diagram of the most representative plane of the domain for droplets at $\mathrm{St}^*=50$.}
    \label{fig:voronoi_St50}
\end{figure}

To more accurately analyze clustering, we quantitatively evaluate it by the standard deviation of ${V}/{\langle V \rangle }$ distribution. $\mathrm{std\left(V/\langle V \rangle \right)}$ is shown in \Cref{fig:std_V}. In Voronoi analyses of particle-laden flows, the random Poisson process (RPP) is commonly used as the baseline. For three-dimensional RPP, $\mathrm{std}(V/\langle V\rangle)$ is approximately 0.42 \citep{kumarPropertiesThreedimensionalPoissonvoronoi1992}. As shown in \Cref{fig:std_V}, $\mathrm{std\left(V/\langle V \rangle \right)}$ for non-inertial droplets is very close to the value for three-dimensional RPP, indicating negligible clustering, while the strongest clustering is observed at $\mathrm{St}^*=10$. Droplet deformation further modulates the clustering behavior. {To achieve the same level of clustering, deformed droplets require a higher Stokes number.} If $\mathrm{St}^*$ is fixed, then for droplets with $\mathrm{St}^*\le1$, deformation reduces the strength of clustering almost monotonically with increasing Weber number, whereas for droplets at $\mathrm{St}^*=50$, clustering is monotonically enhanced. At $\mathrm{St}^*=10$, the effect of deformation is non-monotonic. The clustering is enhanced by deformation at $\mathrm{We}^*=4$, but is suppressed at $\mathrm{We}^*=9$ compared to $\mathrm{We}^*=4$. The Weber number dependency for droplets at $\mathrm{St}^*=50$ is further visually corroborated by the two-dimensional Voronoi diagram in \Cref{fig:voronoi_St50}, which are extracted from a thin slice at the most representative plane in the domain. The representative slice is selected from 33 equally spaced slices, such that its standard deviation of the normalized Voronoi cell area is the closest to the value in \Cref{fig:std_V}.

These results are consistent with our previous analysis in \Cref{sec:cos}. Overall, deformation increases the drag coefficient, thereby reducing the droplet relaxation time and shifting the Stokes number toward lower values. For non-inertial droplets, which already behave as tracer particles, the effect of deformation is negligible. For weakly-inertial droplets, $\langle\cos\theta_{\mathrm{rel},n}\rangle$ is significantly reduced compared to non-inertial droplets, indicating that they are less effectively expelled by centrifugal forces. The reduction in Stokes number caused by deformation then enhances their centrifugal expulsion, increasing the outward normal velocity component and driving them further from vortices, which consequently weakens the preferential concentration. For strongly-inertial droplets, the increased drag due to deformation allows them to respond more actively to the swirling motion of vortices. This is reflected in the less negative tangential direction cosine, indicating a reduced lag behind the swirling motion of vortices. Consequently, these droplets are more readily entrained into the vortex periphery, resulting in an enhanced preferential concentration.

It is interesting to consider the implications of this on the application of spray and droplet combustion. Concentrating droplets closely together tends to decrease the rate of evaporation, and therefore combustion of fuel droplets. However, shearing droplets tends to increase the rate of evaporation, leading to a confounding effect. Furthermore, recent work has also demonstrated that droplet evaporation and combustion rates of deformed droplets differ from those of spherical ones \citep{mashayekDynamicsEvaporatingDrops2001,toniniExactSolutionMass2013,setiyaQuasisteadyEvaporationDeformable2023,setiyaCombustionEvaporationDeformable2023,boydSimulationModelingVaporization2024} which further complicates predictions of this phenomena.

\subsection{{Comparison between steady and unsteady deformation models}}\label{sec:TABunsteady}
To assess the importance of unsteady deformation, we compare the full time-dependent TAB model used throughout \Cref{sec:StWe} with a steady-state approximation. The unsteady model advances the distortion $y$ by evaluating the analytical solution of \Cref{TAB} at the discrete simulation timestep $\Delta t$, thereby preserving the memory of past deformation and transient shape oscillations. The steady model instead uses the long-time limit of the same solution, obtained by taking $t\to\infty$, which eliminates the transient contributions. In this subsection we examine the droplet statistics of the mean Stokes number, Lagrangian time scale, mean-square velocity, droplet dispersion coefficient, and the PDF of Voronoi cell volumes for both models across different droplet inertia at each $\langle\mathrm{We}\rangle^*=9$. The goal is to determine whether the transient behavior predicted by the full TAB model produces a statistically distinguishable effect on dispersion relative to the steady approximation.

\begin{figure}[H]
    \centering
    \begin{subfigure}[b]{0.45\linewidth}
        \centering
        \includegraphics[width=\linewidth]{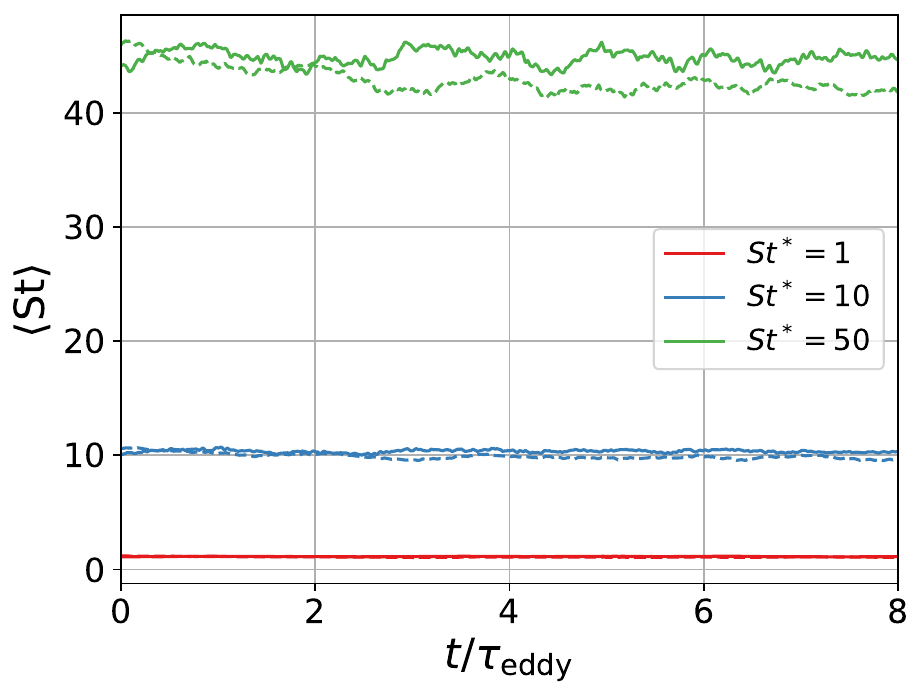}
        \caption{Mean Stokes number of droplets.}
        \label{fig:St_drag_sub}
    \end{subfigure}
    \hfill
    \begin{subfigure}[b]{0.45\linewidth}
        \centering
        \includegraphics[width=\linewidth]{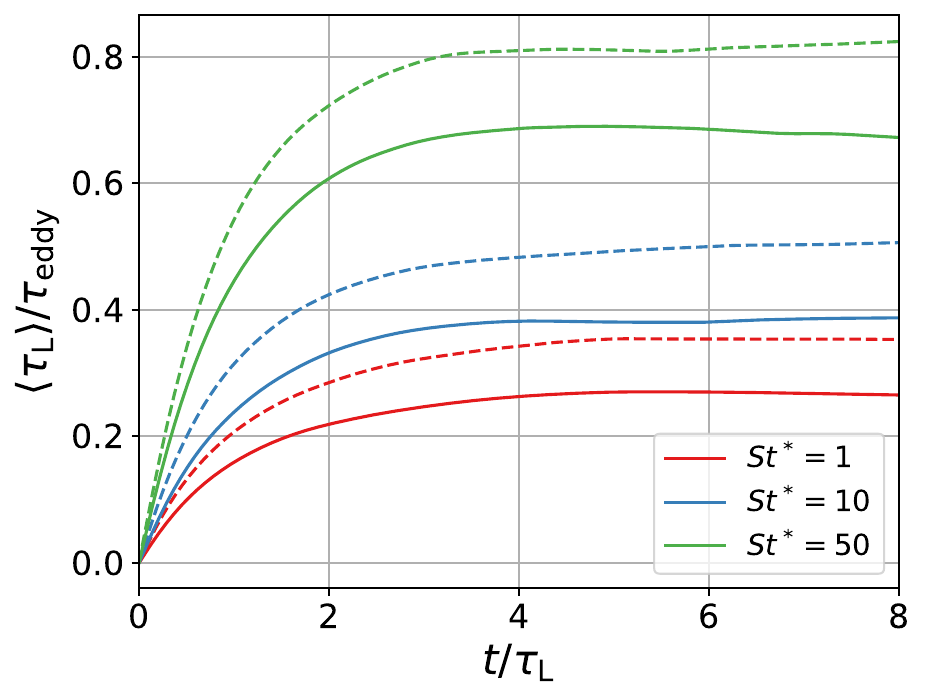}
        \caption{Droplet Lagrangian time scale.}
        \label{fig:tL_drag_sub}
    \end{subfigure}
    \par\medskip
    \begin{subfigure}[b]{0.45\linewidth}
        \centering
        \includegraphics[width=\linewidth]{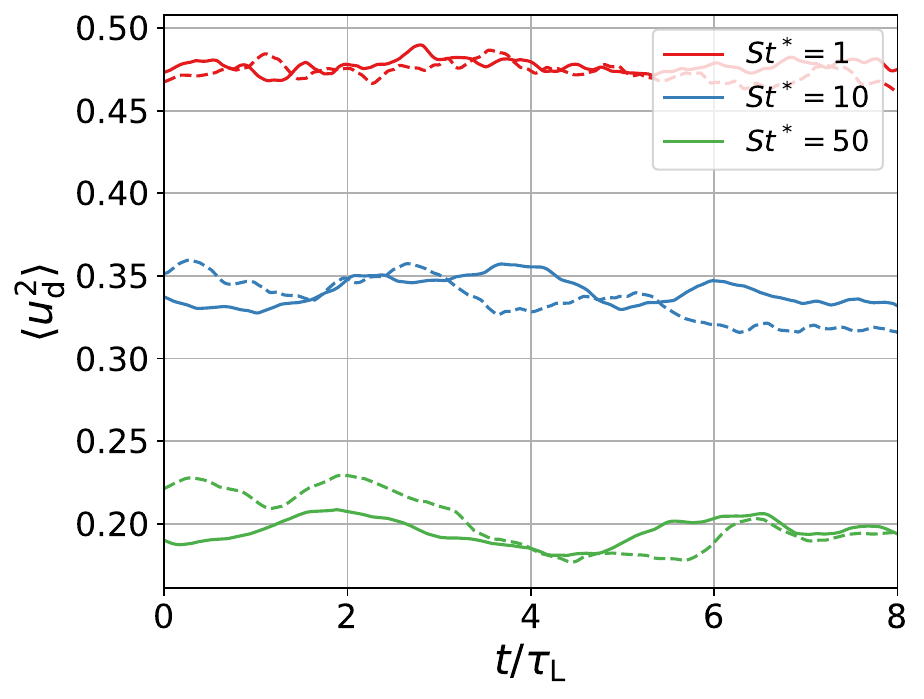}
        \caption{Mean-square droplet velocity.}
        \label{fig:u2_drag_sub}
    \end{subfigure}
    \hfill
    \begin{subfigure}[b]{0.45\linewidth}
        \centering
        \includegraphics[width=\linewidth]{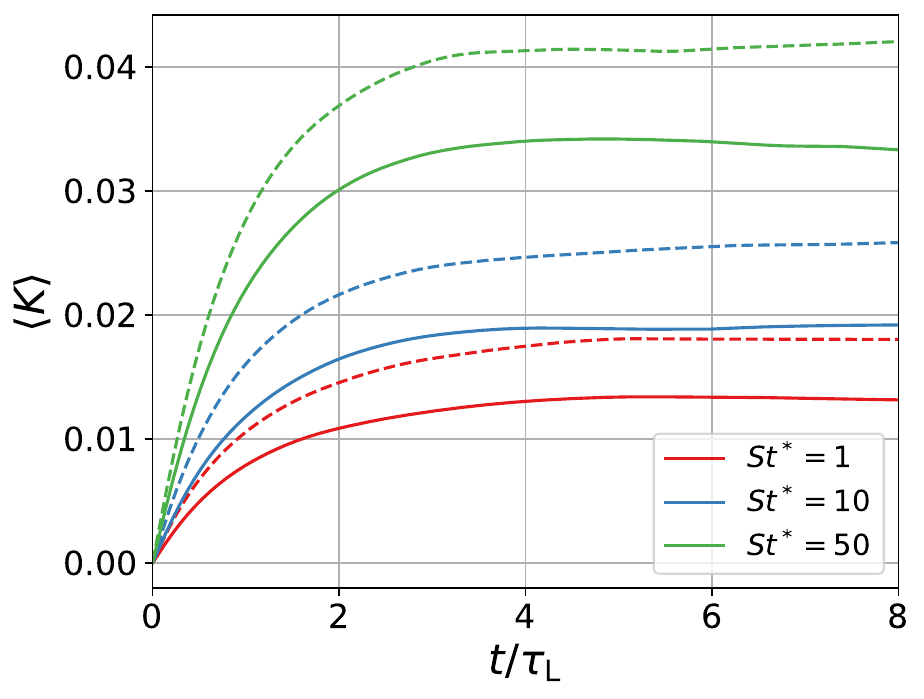}
        \caption{Droplet dispersion coefficient.}
        \label{fig:K_drag_sub}
    \end{subfigure}
    \caption{Comparison of droplet statistics between unsteady and steady deformation models. Solid lines correspond to the unsteady model, while dashed lines correspond to the steady model.}
    \label{fig:all_drag}
\end{figure}

\begin{figure}[H]
    \centering
    \begin{subfigure}{0.32\linewidth}
        \centering
        \includegraphics[width=\linewidth]{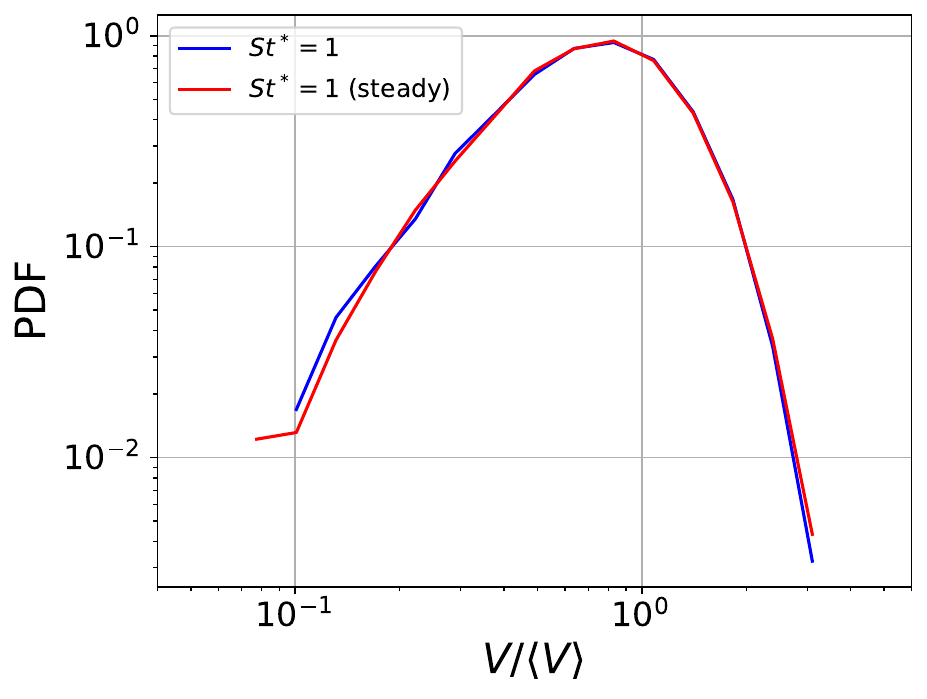}
        \caption{$\mathrm{St}^*=1$}
        \label{fig:pdf_St01We09}
    \end{subfigure}
    \hfill
    \begin{subfigure}{0.32\linewidth}
        \centering
        \includegraphics[width=\linewidth]{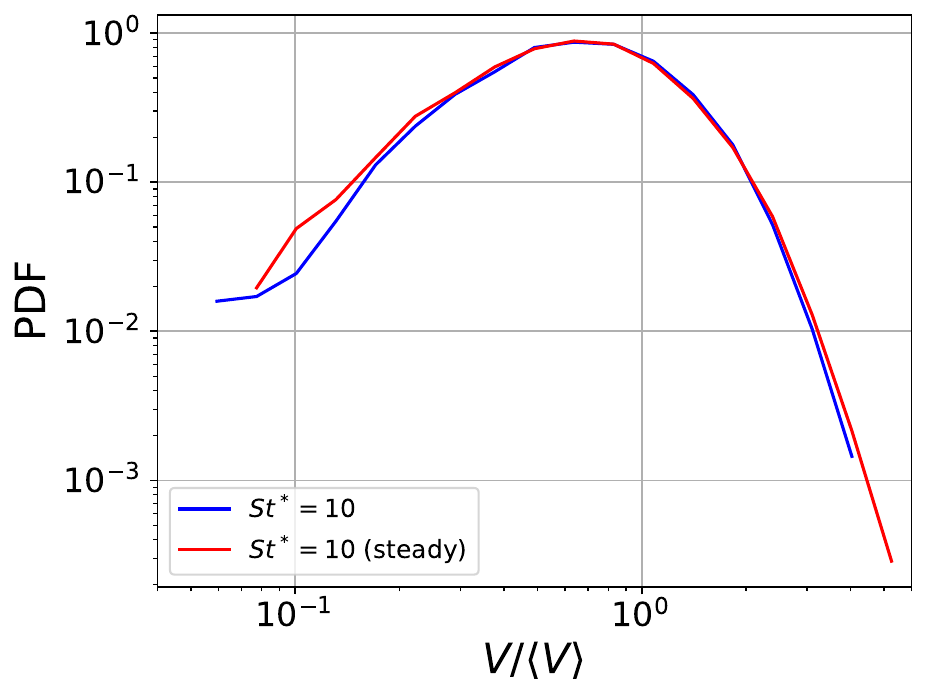}
        \caption{$\mathrm{St}^*=10$}
        \label{fig:pdf_St10We09}
    \end{subfigure}
    \hfill
    \begin{subfigure}{0.32\linewidth}
        \centering
        \includegraphics[width=\linewidth]{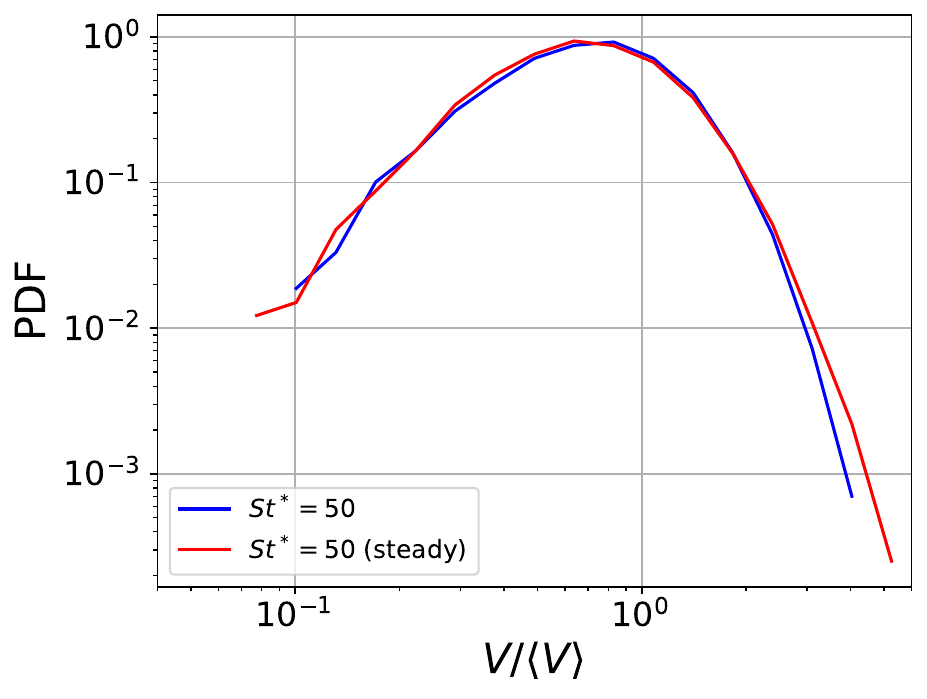}
        \caption{$\mathrm{St}^*=50$}
        \label{fig:pdf_St50We09}
    \end{subfigure}
    \caption{Probability density function of Voronoi cell volume for deformed droplets at $\mathrm{We}^*=9$.}
    \label{fig:pdf_We09}
\end{figure}

In \Cref{fig:St_drag_sub} and \Cref{fig:u2_drag_sub}, we observe that $\langle \mathrm{St} \rangle $ and $\langle u^2_\mathrm{d} \rangle $ are insensitive to unsteady/steady deformation models. This can be understood by the fact that the mean Stokes number is primarily determined by the time-averaged drag coefficient, therefore the transient shape oscillation has negligible effect on $\langle \mathrm{St} \rangle $. Similarly, the mean-square velocity is primarily governed by the response to the mean drag force, and therefore the effect of unsteady shape oscillations on $\langle u^2_\mathrm{d} \rangle $ is largely averaged out over time. The PDF of Voronoi cell volumes plotted in \Cref{fig:pdf_We09} also suggests that the clustering of droplets is unaffected by the unsteady/steady deformation models, as the two PDFs are nearly identical, indicating that preferential concentration is primarily determined by the mean Stokes number. However, the Lagrangian time scale $\tau_\mathrm{L}$ and droplet dispersion coefficient $K$ shown in \Cref{fig:tL_drag_sub} and \Cref{fig:K_drag_sub} are significantly affected by the choice of the two models. The steady model predicts a much larger $\langle \tau_\mathrm{L} \rangle $ and $\langle K \rangle $ than the unsteady model. This discrepancy arises because the transient shape oscillations captured by the unsteady model have introduced additional fluctuations in the drag force, enhancing the decorrelation of droplet velocity, and consequently reducing $\langle \tau_\mathrm{L} \rangle $. Then, since $\langle u^2_\mathrm{d} \rangle $ remains identical between the two deformation models, the reduction of $\langle \tau_\mathrm{L} \rangle $ directly leads to the reduction of $\langle K \rangle $ according to Taylor's dispersion theory.

\section{Conclusion}
\label{sec:conclud}
This study reveals the coupled effect of droplet deformation and inertia on droplet dispersion and clustering in homogeneous isotropic turbulence via direct numerical simulation with Lagrangian particle tracking method. By systematically varying the Stokes number and the Weber number of droplets, we quantify droplet dispersion through the classical Taylor's dispersion theory framework, a local vortex coordinate decomposition analysis, and Voronoi analysis. The principal findings are summarized as follows.

Deformation increases the drag coefficient and therefore shortens the droplet relaxation time, shifting the instantaneous Stokes number toward smaller values. This effect is substantial for strongly-inertial droplets but negligible for non-inertial droplets. For droplet dispersion, the effect of deformation is dependent on the inertia regime. For non-inertial droplets, deformation enhances dispersion, while for strongly-inertial droplets, dispersion is suppressed. The change of droplet dispersion coefficient due to deformation is much smaller than that of droplet Lagrangian time scale. This is because for strongly-inertial droplets, deformation reduces droplet Lagrangian time scale but simultaneously increases the mean-square velocity, making the dispersion coefficient less sensitive to changes in the Weber number.

Analysis of direction cosines in local vortex coordinates and Voronoi analysis reveals that the effect of deformation on clustering is inertia regime dependent as well. For non-inertial droplets, deformation enhances the centrifugal expulsion effect of vortices, driving droplets further away from vortices and consequently weakening clustering. For strongly-inertial droplets, deformation allows them to respond more actively to the swirling motion of vortices, which makes them more readily entrained into the vortex periphery and results in an enhanced preferential concentration.

A comparison of the full time-dependent TAB model with its steady-state limit further shows that the Lagrangian time scale and droplet dispersion coefficient are sensitive to the unsteady shape dynamics retained by the model, whereas the mean Stokes number, mean-square velocity, and Voronoi clustering patterns are not. This indicates that the transient oscillations captured by the TAB model primarily affect temporal correlation statistics, while the energy-related properties and the spatial distribution governed by the time-averaged drag are less sensitive to the unsteady deformation dynamics.

Overall, these results demonstrate that droplet deformation fundamentally alters the dispersion and clustering of droplets in turbulence through a combined modulation of drag, inertia, and vortex interaction. Accounting for deformation is therefore essential for predictive spray modeling in turbulent combustion applications. Future work should investigate how these phenomena interplay with droplet evaporation and heat release.

\section*{Acknowledgments}
This work was performed as part of the Space Ignite Center for Advanced Research-Education in Combustion (SPARC) [NASA Award 80NSSC24M0173]. Parts of this work were completed on \emph{Hyak}, UW's high performance computing cluster. This work used the NSF ACES supercomputer at Texas A\&M High Performance Research Computing through allocation MCH240016 from the Advanced Cyberinfrastructure Coordination Ecosystem: Services \& Support (ACCESS) program, which is supported by U.~S. National Science Foundation grants \#2138259, \#2138286, \#2138307, \#2137603, and \#2138296. 

\bibliographystyle{elsarticle-num-names}
\bibliography{cas-refs}

\end{document}